\documentclass[a4paper,useAMS,usenatbib]{mnras}

\usepackage{graphicx}
\usepackage{setspace}
\usepackage{natbib}
\usepackage{color}
\usepackage{amsmath,amssymb}
\usepackage{times}
\def\Ms{{M_{\rm s}}}

\title[The biggest Splash]{The biggest Splash}

\author[Belokurov et al]{Vasily
  Belokurov$^{1}$\thanks{E-mail:vasily@ast.cam.ac.uk},
  Jason L. Sanders$^{1}$, Azadeh Fattahi$^{2}$, Martin C. Smith$^{3}$, Alis J. Deason$^{2}$, \newauthor N. Wyn Evans$^{1}$ and Robert J. J. Grand$^{4,5,6}$\\ $^{1}$Institute of Astronomy, Madingley Rd,
  Cambridge, CB3 0HA\\
  $^{2}$Institute for Computational Cosmology, Department of Physics, University of Durham, South Road, Durham DH1 3LE, UK\\
  $^{3}$ Key Laboratory for Research in Galaxies and Cosmology, Shanghai Astronomical Observatory, Chinese Academy of Sciences, 80 Nandan Road, \\
  Shanghai 200030, China\\
  $^4$ Heidelberger Institut f\"ur Theoretische Studien, Schlo{\ss}-Wolfsbrunnenweg 35, D-69118 Heidelberg, Germany\\
  $^5$Zentrum f\"ur Astronomie der Universit\"at Heidelberg, Astronomisches Recheninstitut, M\"onchhofstr. 12-14, D-69120 Heidelberg, Germany\\
  $^6$Max-Planck-Institut f\"ur Astrophysik, Karl-Schwarzschild-Str. 1, D-85748, Garching, Germany}

\hypersetup{draft}
\begin{document}



\maketitle

\label{firstpage}

\begin{abstract}
Using a large sample of bright nearby stars with accurate {\it Gaia}
Data Release 2 astrometry and auxiliary spectroscopy we map out the
properties of the principle Galactic components such as the ``thin''
and ``thick'' discs and the halo. We show that in the Solar
neighborhood, there exists a large population of metal-rich
([Fe/H]$>-0.7$) stars on highly eccentric orbits. By studying the
evolution of elemental abundances, kinematics and stellar ages in the
plane of azimuthal velocity $v_{\phi}$ and metallicity [Fe/H], we
demonstrate that this metal-rich halo-like component, which we dub the
{\it Splash}, is linked to the $\alpha$-rich (or ``thick'')
disc. Splash stars have little to no angular momentum and many are on
retrograde orbits. They are predominantly old, but not as old as the
stars deposited into the Milky Way in the last major merger. We argue,
in agreement with several recent studies, that the Splash stars may
have been born in the Milky Way's proto-disc prior to the massive
ancient accretion event which drastically altered their orbits. We can
not, however, rule out other (alternative) formation channels. Taking
advantage of the causal connection between the merger and the Splash,
we put constraints of the epoch of the last massive accretion event to
have finished 9.5 Gyr ago. The link between the local metal-rich and
metal-poor retrograde stars is confirmed using a large suite of
cutting-edge numerical simulations of the Milky Way's formation.
\end{abstract}


\section{Introduction}

Until recently, chronology of the Milky Way's assembly had been mostly
hypothesized, but now it is possible to {\it determine} the formation
history of our Galaxy thanks to data from ESA's {\it Gaia} space
observatory \citep[][]{Prusti2016}. By combining {\it Gaia}'s
astrometry with auxiliary spectro-photometry, ages can be measured for
increasingly large samples of stars \citep[see e.g.][]{SandersDas2018}.
Thus we can now hope to establish the order in which the principal
Milky Way components - such as thick and thin discs, bar, bulge and
halo (stellar and dark) - were created. Of these, the chronology of
the stellar halo is the simplest to piece together, as the remains of
the individual building blocks can stay observable almost
indefinitely \citep[][]{Helmi2000}. Within the $\Lambda$CDM Cosmology,
the bulk of the stellar halo of a Milky Way-sized galaxy is predicted
to be dominated by a small number of massive early-accreted dwarfs
\citep[e.g.][]{BJ2005,Robertson2005,Font2006,DeLucia2008}. In
connection with this, \citet{Deason2013} put forward the hypothesis that
the rapid transition in the Galactic stellar halo structural
properties at the so-called break radius of 20-30 kpc
\cite[see][]{Watkins2009,Deason2011,Sesar2011} is due to the
apo-centre pile-up of the tidal debris from a small number of
significant mergers early in the life of the Milky Way.

The \citet{Deason2013} hypothesis has recently been tested with the
{\it Gaia} data. Using {\it Gaia} Data Release 1 (DR1) and the
re-calibrated SDSS astrometric catalogues, \citet{Sausage} and
\citet{ActionHalo} point out the local dominance of a relatively
metal-rich, mildly prograde and highly radially-anisotropic stellar
halo component. In fact, the anisotropy is so strong that the local
velocity distribution appears highly stretched in the radial
direction, taking a cigar-like or sausage-like shape. Comparing the
{\it Gaia}-DR1-SDSS data to cosmological zoom-in simulations,
\citet{Sausage} argue that only massive ($\sim10^{11} M_{\odot}$)
early (8-11 Gyr ago) mergers can deliver large amounts of stellar
material with high orbital anisotropy into the Solar
neighborhood. Similarly, \citet{Haywood2018} use {\it Gaia}~Data
Release 2 (DR2) to contend that the debris from a single significant
merger must be the cause of the prominence of stars with high
tangential velocity and intermediate metallicity, i.e. [Fe/H]$<$-1 in
the local {\it Gaia} samples.

Subsequently, the properties of this ancient Gaia-Sausage (GS) merger
(sometimes also referred to as Gaia Enceladus) have been further
constrained using the {\it Gaia} DR2 data (GDR2). For example,
\citet{Helmi2018} use GDR2 astrometry and APOGEE spectroscopy to show
that stars on highly eccentric orbits follow a characteristic track in
the [$\alpha$/Fe] and [Fe/H] plane, thus yielding a constraint on the
mass of the progenitor and the epoch of interaction. From the
isochronal age of the GS stars \citet{Helmi2018} estimate the merger
to have happened $\sim$10 Gyr ago. \citet{Mackereth2019} also look at
the APOGEE data, highlighting a tight connection between the chemistry
and the orbital properties of the Gaia Sausage stars. By comparing the
observed stellar halo chemo-dynamics to that emerging from the EAGLE
simulations suite, \citet{Mackereth2019} gauge the stellar mass of the
progenitor to be $<10^9 M_{\odot}$ and the accretion redshift
$z\sim1.5$.

While these original studies presented a broad-brush interpretation of
the local stellar halo properties, in the last year there has been
plenty of follow-up work that has filled in the missing details.  For
example, \citet{Vincenzo2019} model the chemical evolution of the GS
candidate stars as selected using the APOGEE and GDR2 data-sets, and
find a total stellar mass at the time of the merger of order of
$\sim10^9 M_{\odot}$, with the bulk of the dwarf's stars forming more
than 10 Gyr ago. The presence of the GS merger debris was also
revealed through a careful analysis of the local halo sample
\citep[][]{Necib2019} and traced far beyond the Solar neighborhood
\citep[][]{Lancaster2019,Bird2019}. \citet{PileUp} connect the orbital
properties of the local Main Sequence stars and the distant Blue
Horizontal Branch stars and show that the two samples have very
similar apo-centric distance, thus establishing that the local and the
far-field halo tracers likely share a common progenitor. Using
independent sets of halo tracers, \citet{Wegg2019} and
\citet{Tian2019} confirm that the GS debris has a small but
statistically significant prograde rotation in agreement with
\citet{SlightSpin, Sausage, ActionHalo}.  \citet{Sequoia} present
chemical and kinematic evidence for two (perhaps related) accretion
events: the nearly radial GS merger \citep[see][]{Sausage} and the
highly retrograde Sequoia \citep[see also][]{Shards,Matsuno2019}. The
structural properties of the GS debris cloud were mapped out in
\citet{Iorio2019}, who demonstrate that the bulk of the RR Lyrae stars
within 20-30 kpc were probably once part of the GS progenitor
galaxy. According to \citet{Iorio2019}, the inner halo RR Lyrae
distribution is triaxial, with the major axis almost in the Magellanic
Cloud orbital plane. Curiously, at either end of this triaxial
ellipsoid there reside two previously known halo sub-structures: the
Virgo Overdensity and the Hercules-Aquila Cloud. \citet{Simion2019}
make a compelling case that both VO and HAC are likely unmixed
portions of the GS tidal debris.

Globular clusters (GCs) can provide a unique window into the early
accretion history of the Galaxy. Several studies took advantage of the
fast growing number of GCs with accurate ages and kinematics to
produce independent constraints on the early merger events in the
MW. In a pioneering study, \citet{Kruijssen2019} claim to have
identified three significant early mergers, although note that their
assignment of individual GCs to particular events does not use any
orbital information. \citet{SausageGCs} complemented the sample
provided by \citet{Kruijssen2019} with the GC orbital properties from
\citet{GaiaSatPM} to discover a large group of clusters likely
originating from the GS dwarf galaxy \citep[see also][]{Piatti2019},
thus confirming the high mass of the progenitor. \citet{Sequoia}
refine the Gaia Sausage membership probability for the Galactic GCs
\citep[see also][]{Massari2019} and estimate the dwarf's total stellar
mass of up to $5\times10^9 M_{\odot}$ prior to its disruption in the
MW. Note that most recently, \citet{SHMass} came up with a direct
measurement of the total Galactic stellar halo mass using the {\it
  Gaia} DR2 kinematics. Their value of $\sim10^9 M_{\odot}$ is in good
agreement with the above indirect estimates.

There have also been several attempts to compare the observational
evidence for an early massive merger with the numerical simulations of
the Milky Way's formation. Looking in the Aquarius simulation suite,
\citet{Emma2019} see a variety of accretion pathways imprinted in the
distinct alpha-iron abundance trends as a function of Galactocentric
radius, albeit without a clear match to the MW observations described
above. \citet{Mackereth2019} identify counterparts of the observed
local high-eccentricity stars in the EAGLE suite of Cosmological
zoom-in simulations. Complementing their study with [Mg/Fe]-[Fe/H]
information, they arrive at the stellar mass of the progenitor galaxy
of the order of $10^9 M_{\odot}$. \citet{Fattahi2019} take a different
approach, attempting to reconstruct the details of the merger by
comparing the observed properties of the local stellar halo to those
of the simulated Auriga galaxies \citep[see][for
  details]{Grand2017}. Following \citet{Sausage}, \citet{Fattahi2019}
model the local velocity distribution with a mixture of multi-variate
Gaussians and find that those Auriga halos that contain a dominant
velocity component with an orbital anisotropy $\beta>0.8$ are
typically assembled early, undergoing a merger with SMC/LMC-sized
dwarf galaxy 6-11 Gyr ago.

Note however that, in numerical simulations of galaxy formation,
stellar halos are typically assembled through at least two distinct
channels. In mock Milky Ways, stars at large distances from the
Galactic plane (following a loose definition of a halo) can be either
accreted from smaller satellites \citep[see
  e.g.][]{Johnston1996,Helmi1999,BJ2005,Cooper2010} or born in-situ
\citep[][]{Zolotov2009,McCarthy2012,Tissera2012,Cooper2015}. The
details of the in-situ halo formation differ between the above
studies. In particular, \citet{Zolotov2009} identify the in-situ halo
as stars born very early on in the very inner portions of the
progenitor MWs in a spheroidal configuration. \citet{McCarthy2012}
argue that this mechanism is likely not viable and is probably linked
to over-cooling; they suggest instead that the in-situ halo forms
somewhat later in a proto-disc which is subsequently heated up to
large distances. \citet{Cooper2015} list three different channels of
in-situ stellar halo formation. Curiously, according to
\citet{Cooper2015}, the proto-disc heating provides the smallest
contribution to the in-situ halo, but can dominate in the Solar
neighborhood with a caveat that an unambiguous distinction between the
thick disc and the in-situ halo in this region may be
difficult. Interestingly, they also see in-situ halo star formation
directly from gas particles accreted onto the MW. This can be either
smoothly accreted gas of the gas stripped from gas-rich
satellites. \citet{Cooper2015} caution however that this mode of star
formation may be an artefact of the sub-grid physics prescriptions
used in the simulations.

A search for an in-situ stellar halo component has been on for some
time now. It is however clear from the discussion above that depending
on the formation scenario, the present day in-situ stellar halos can
have drastically different properties. \citet{Zolotov2009} and
\citet{McCarthy2012} agree that the in-situ component ought to be old
and metal-rich, while forecasting rather different kinematics: in one case,
in-situ halos do not have to have any appreciable angular momentum
while in the other, they typically maintain significant net
rotation. On the other hand, if stars can be born in the halo from the
gas stripped from gas-rich dwarfs \citep[as suggested
  by][]{Cooper2015}, at redshift z=0, such an in-situ component will
have properties indistinguishable from that of the accreted halo. Some
authors took the observation by \citet{Carollo2010} that the MW
stellar halo evolves from metal-rich at small Galactocentric radii to
metal-poor further out as evidence for the presence of an in-situ
stellar halo in the inner Galaxy. It is however becoming increasingly
clear that the metal-rich ([Fe/H]$\sim-1.6$) ``inner halo'' identified
by \citet{Carollo2010} is nothing but the tidal debris from the
ancient GS merger, whose extent is largely limited to 20-30 kpc
\citep[see][]{Deason2013,PileUp}.

The question of the existence of the Galactic in-situ stellar halo has
been revived in light of the recent {\it Gaia} data releases. Using
the {\it Gaia} DR1 data augmented with the RAVE and APOGEE
spectroscopy, \citet{Bonaca2017} build a large sample of
high-eccentricity stars with [Fe/H]$>-1$. Using the available chemical
information and by comparing their observations to numerical
simulations, \citet{Bonaca2017} conclude that these locally-observed
metal-rich stars on halo-like orbits have likely been born
in-situ. They estimate that this in-situ component contributes as much
as $50\%$ of the halo stars locally. \citet{Haywood2018},
\citet{DiMatteo2018} and Amarante et al. (2019) use {\it Gaia} DR2
to detect a substantial (again close to $\sim50\%$) population of
high-eccentricity stars with thick disc chemistry, in agreement with
the {\it Gaia} DR1 study of \citet{Bonaca2017}. These authors choose
not to label this component an ``in-situ halo'' but rather a ``heated
thick disc'', perhaps in line with the earlier definition of the
in-situ halo in \citet{Zolotov2009}. Note however that the in-situ
stellar halo is defined as a heated pre-historic disc in the works of
\citet{McCarthy2012} and \citet{Cooper2015}. \citet{Gallart2019}
concentrate on the split Main Sequence discovered by \citet{Babu2018}
and claim that the stars analyzed by
\citet{Bonaca2017,Haywood2018,DiMatteo2018} and earlier by
\citet{Nissen2010,Nissen2011,Schuster2012} are typically as old as the
oldest accreted halo stars and significantly older than the bulk of
the thick disc population, therefore resuscitating the idea of
\citet{Zolotov2009} of the ancient MW spheroid formation.

\begin{figure*}
  \centering
  \includegraphics[width=0.97\textwidth]{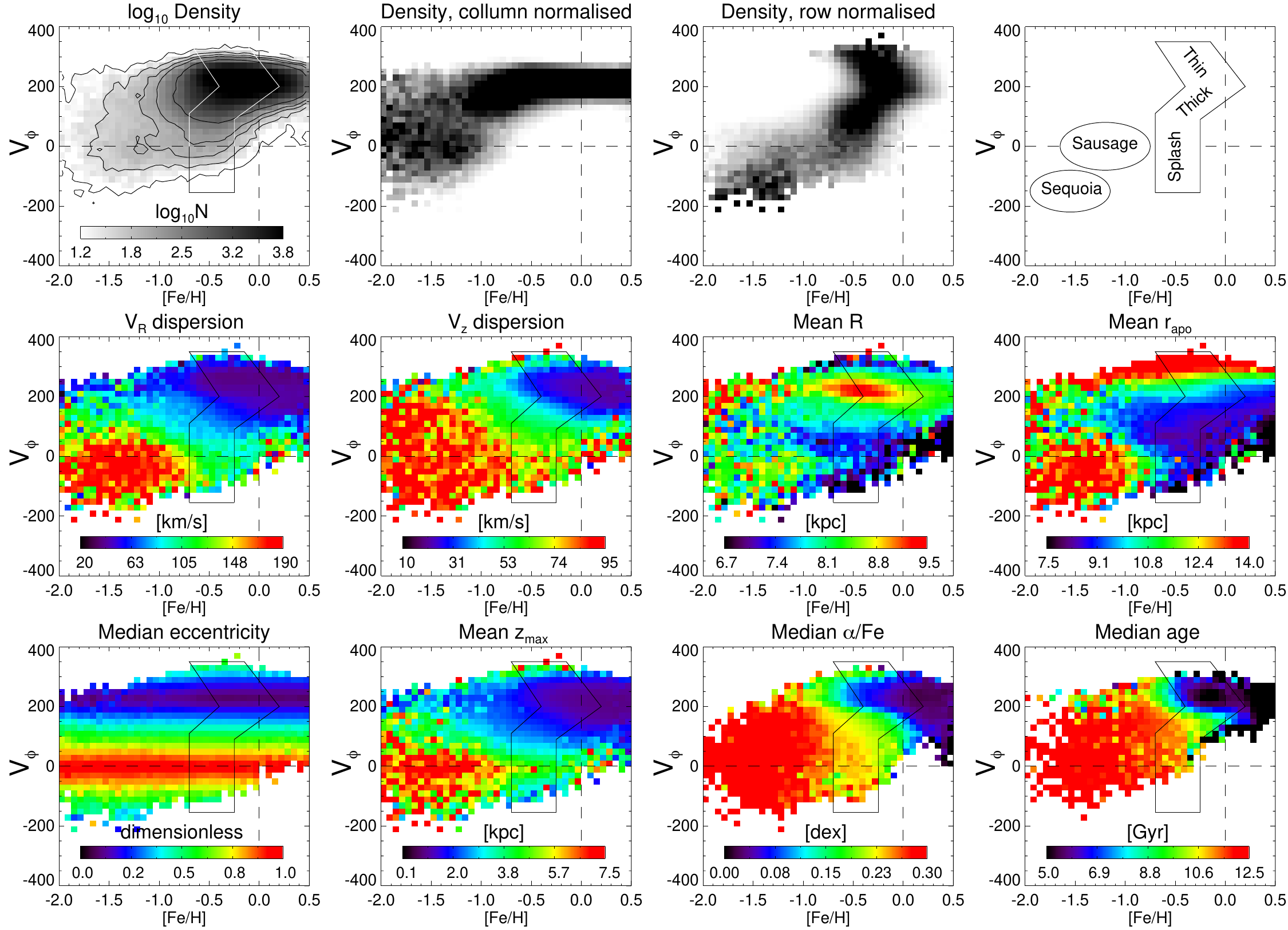}
  \caption[]{Stars with $2<|z| (\mathrm{kpc})<3$ in the plane of
    $v_{\phi}$-[Fe/H]. {\it Top row:} First panel shows the logarithm
    of the stellar density. Second panel gives column-normalized
    stellar density. Note the bi-modality in the $v_{\phi}$
    distribution at $-1.5<$[Fe/H]$<-0.7$, where the high $v_{\phi}$
    population is the Galactic disc, and low $v_{\phi}$ population is
    the Gaia Sausage. Third panel shows the row-normalzed stellar
    density. Note the chevron-like $>$ pattern of the thin and thick
    discs (see text for details) as well as the very metal-rich
    population with low and/or negative $v_{\phi}$. Fourth panel
    presents a schematic summary of the feature seen in the previous
    panels. {\it Middle row:} First panel shows the radial velocity
    dispersion. Note a clear dichotomy between the accreted halo with
    [Fe/H]$<-1$ and $v_{\phi}<100$ km s$^{-1}$ seen in red here and
    the Galactic discs with [Fe/H]$>-1$ and $v_{\phi}>100$ km s$^{-1}$
    seen in blue. The Splash stars (mostly green) with
    $-0.7<$[Fe/H]$-0.2$ and $v_{\phi}<100$ km s$^{-1}$ have radial
    velocity dispersion significantly higher than that of the disc but
    lower than the Gaia Sausage. Second panel gives the vertical
    velocity dispersion. Third: mean instantaneous cylindrical
    radius. Compared to GS stars, the Splash is preferentially
    observed inside of the Solar radius. Fourth panel: Mean
    apo-centric distance. Note a clear difference between the GS and
    Splash stars. {\it Bottom row:} First panels gives the median
    eccentricity. All stars with $v_{\phi}<100$ km s$^{-1}$ have
    $e>0.5$.  Second: Mean $z_{\rm max}$. Note a transition to much
    larger heights at $v_{\phi}\sim100$ km s$^{-1}$ and
    $-0.7<$[Fe/H]$<-0.2$. Third: median $\alpha$-abundance. The Splash
    $[\alpha/\mathrm{Fe}]$ ratios are the highest at the corresponding
    metallicity and overlap with low-$v_{\phi}$ tail of the thick
    disc. Fourth: median stellar age. The Splash stars are slightly
    younger than those in GS and overlap with the old age (and low
    $v_{\phi}$) tail of the thick disc.}
   \label{fig:grid}
\end{figure*}

In this Paper, we study chemo-chronological properties of a large
sample of stars on halo-like orbits, i.e. those with low or no angular
momentum. Our focus is on the metal-rich, more precisely
$-1<$[Fe/H]$<0$ portion of this population, which we demonstrate to be
clearly distinct from the accreted (and typically more metal-poor)
stars of the Gaia Sausage. We show that at the same metallicity, there
exists a smooth transition from this low-angular momentum population
to the rapidly rotating ``thin'' and ``thick'' discs. We introduce our
dataset in Section~\ref{sec:sample} and highlight the differences
between the metal-poor and metal-rich halo components in
Section~\ref{sec:sauspl}. We extend out analysis beyond the Solar
neighborhood in Section~\ref{sec:beyond}. Finally, in
Section~\ref{sec:disc}, the observations are compared to numerical
simulations of the Milky Way's formation. We also invoke virial
theorem to provide an analytic estimate of the mass of the massive
perturber based on the energetics of the encounter.

\section{A glimpse of the Galaxy in 9D}
\label{sec:sample}

Our primary dataset is based on the large sample of nearby stars with
accurate {\it Gaia} DR2 \citep[][]{GaiaDR2} astrometry and auxiliary
spectroscopy, augmented with 2MASS \citep[][]{2MASS}, {\it Gaia} and
Pan-STARRS \citep[][]{PS1} photometry as produced by
\citet{SandersDas2018}. In particular, spectroscopic information from
the APOGEE \citep[][]{Majewski2017, Garcia2016} DR14
\citep[][]{APOGEEDR14}, LAMOST DR3 \citep[][]{Cui2012, Zhao2012}, RAVE
\citep[][]{RAVE} DR5 \citep[][]{RAVEDR5}, {\it Gaia}-ESO DR3
\citep[][]{GES}, GALAH \citep[][]{DeSilva2015, Martell2017} DR2
\citep[][]{Buder2018} and SEGUE \citep[][]{Yanny2009} surveys is
included. The data were compared to a grid of PARSEC isochrone models
\citep{Bressan2012,Chen2014,Tang2014,Chen2015} in a probabilistic
sense. For giant stars with carbon and nitrogen measurements, constraints on the mass were also used using the method of \cite{DasSanders2019}. Not only do \citet{SandersDas2018} use the extant spectroscopic
information to constrain their distance estimates and measure stellar
ages, they also re-calculate stellar atmosphere parameters, thus
unifying the spectroscopic output across all surveys considered. There
are three features of the \cite{SandersDas2018} method worth
mentioning: (i) the maximum age isochrone considered is
$12.6\,\mathrm{Gyr}$, (ii) there is a weak prior coupling 3D location
in the Galaxy, age and metallicity, and (iii) no offset was applied to
the Gaia parallaxes as its magnitude for the sample was, at the time,
not well characterised. The impact of the latter of these is somewhat
alleviated by also using spectro-photometric information, although
recent studies \citep[e.g.][]{Schonrich2019} have shown the offset for
bright stars, such as those in the spectroscopic overlap, is
significantly greater than for the quasar sample from
\cite{Lindegren2018}. This will be discussed in more detail later.

Given that there exist substantial overlaps between the above surveys,
we restrict the sample to the subset of stars with unique {\it Gaia}
\texttt{source\_id}. From the \citet{SandersDas2018} catalogue of
4,906,746 (3,706,733) entries including (excluding) duplicates we
select unique objects with a small azimuthal velocity error
$\sigma_{v_{\phi}}<20$ km s$^{-1}$ and low metallicity uncertainty
$\sigma_{{\rm [Fe/H]}}<0.15$ as well as accurate parallax measurement
$|\varpi/\sigma_{\varpi}|>5$. The above cuts leave the total of
2,655,034 stars available. Additionally, when considering
alpha-abundance, we also require $\sigma_{\alpha/{\rm Fe}}<0.1$, which
reduces the sample down to 815,930. Our sample is diminished from
2,655,034 down to 514,286 when the stellar ages are analyzed, in
particular we set $\sigma_{\rm age}<2\,\mathrm{Gyr}$ and only include
upper Main Sequence and Turn-off stars, i.e. those with $5700 {\rm K}
<T_{\rm eff} < 8300 {\rm K}$ and $3.5 < \log g < 4.2$. The stellar
velocity components as well as the orbital properties discussed below
are those reported in the catalog of \citet{SandersDas2018}.

Figure~\ref{fig:grid} shows the distribution of stars in the space of
(cylindrical) azimuthal velocity $v_{\phi}$ and metallicity [Fe/H] (as
derived by \citet{SandersDas2018}). Only stars between 0.5 and 3 kpc
away from the Galactic plane are shown, which reduces the sample size
by a factor of $\sim$3. As is clear from the first panel in this grid
(top left), there exists a small yet noticeable population of
metal-rich ([Fe/H]$>-1$) stars with no appreciable angular momentum,
i.e. $v_{\phi} < 100$ km s$^{-1}$. Strikingly, many of these stars are
on radial ($v_{\phi}\sim 0$ km s$^{-1}$) and some are on retrograde
($v_{\phi}<0$ km s$^{-1}$) orbits in agreement with previous studies
\citep[in particular][]{Bonaca2017,Haywood2018,DiMatteo2018}. The next
panel in the top row presents the column-normalized density which
allows one to track the peak of $v_{\phi}$ distribution as a function
of [Fe/H]. Here, at the intermediate metallicity of [Fe/H]$\sim-1.3$
two distinct modes can be seen, one with $v_{\phi}\sim 150$ km
s$^{-1}$ corresponding to the thick disc and one with $v_{\phi}\sim0$
km s$^{-1}$ corresponding to the stellar halo, more precisely to its
dominant local component, the Gaia Sausage \citep[see][]{Sausage}. The
third panel in the top row of Figure~\ref{fig:grid} shows the
row-normalised density in the space of $v_{\phi}$ and [Fe/H], which
reveals the typical metallicity for the given range of azimuthal
velocity. Starting from the highest valiues of $v_{\phi}$ and going
down, a bifurcated chevron-like $>$-sign feature is discernible,
corresponding to the ``thin'' (upper portion of $>$) and the ``thick''
(lower portion of $>$) discs \citep[see also][]{lee2011}. The ``thin''
disc stars show negative $v_{\phi}$ gradient as a function of
metallicity. This behavior arises in the Solar neighborhood for the
same reason as the so-called asymmetric drift. Stars with higher
$v_{\phi}$ are those for which their motion on the epicylcic ellipse
adds to the motion of the guiding centre --- such stars have to come
from Galactocentric radii larger than Solar. Given the negative
metallicity gradient in the Galactic disc, these faster stars are also
more metal-poor. The ``thick'' disc shows an inverted
$v_{\phi}$-metallicity gradient. This is a manifestation of the
so-called Simpson-Yule paradox and is the consequence of the
inside-out disc formation
\citep[see][]{Schonrich2017,Kawata2018,Minchev2019}.

\begin{figure*}
  \centering
  \includegraphics[width=0.98\textwidth]{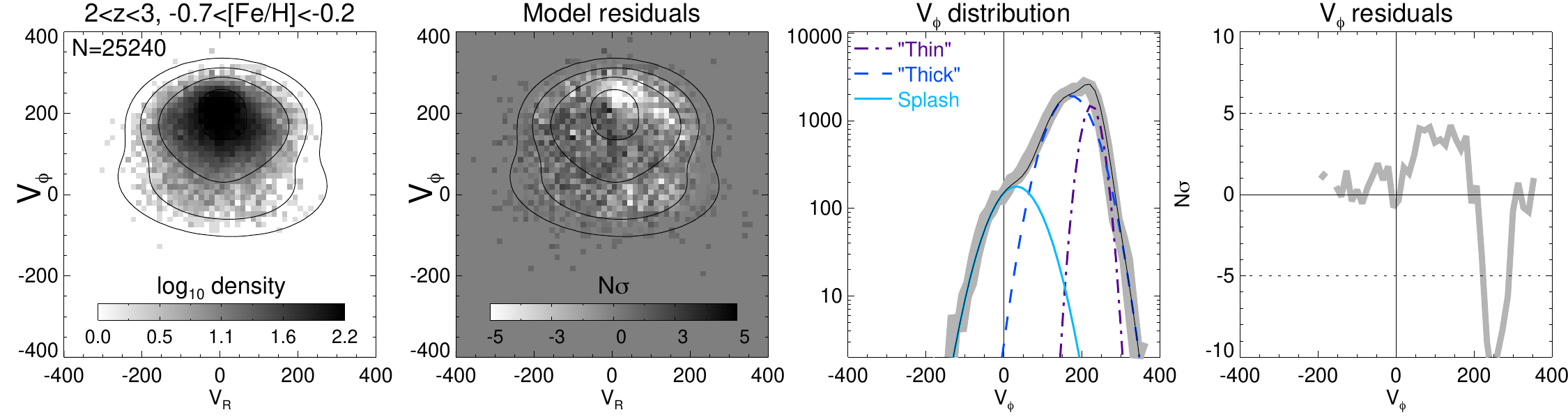}
  \includegraphics[width=0.98\textwidth]{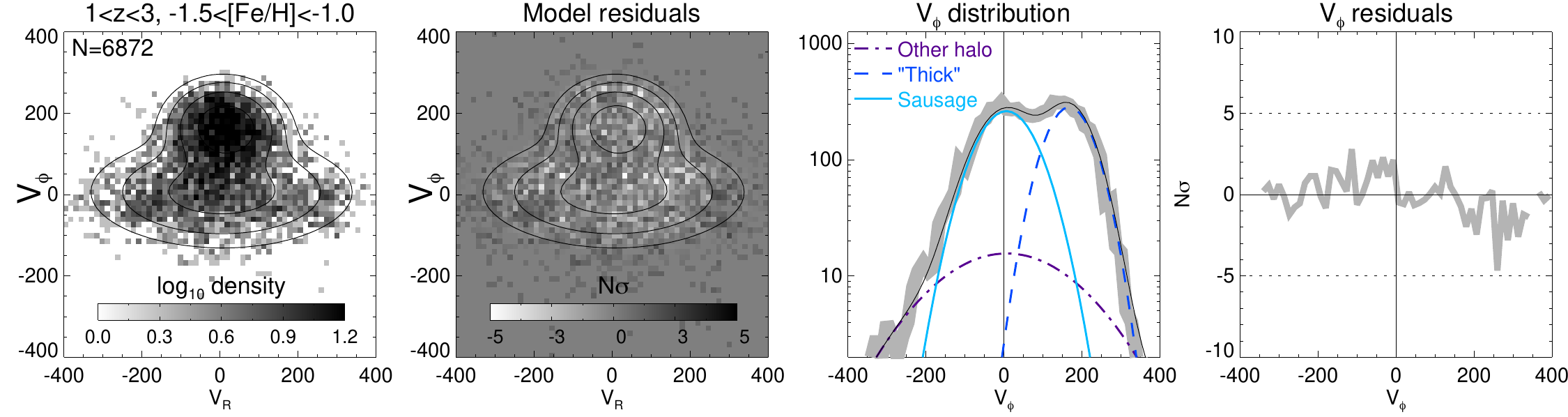}
  \caption[]{The Splash (top) and the Sausage (bottom) in velocity
    space. {\it First panel:} Greyscale density shows the observed
    distribution of stars in $v_{\phi}-v_{\rm R}$ space. Black
    contours show the 3-component Gaussian mixture model. {\it Second
      panel:} Distribution of the model residuals reported as numbers
    of Poisson sigmas. {\it Third panel:} Histogram of the observed
    $v_{\phi}$ velocities (thick grey line) together with model (solid
    thin black line) and three individual components (explained in the
    legend). {\it Fourth panel:} Distribution of the model residuals
    as a function of $v_{\phi}$. Dotted lines show $\pm5\sigma$. Note
    that in the Top row, unsurprisingly, our over-simplified model
    does not describe the disc properties adequately (high residuals
    for $v_{\phi}>100$ km s$^{-1}$. Nonetheless, around
    $v_{\phi}\sim0$ km s$^{-1}$ the model and the data agree well
    allowing us to estimate the properties of the Splash population.}
   \label{fig:vel_fit}
\end{figure*}

At $v_{\phi}\sim0$ km s$^{-1}$ the density runs without clear
interruptions across almost the entire range of metallicities. Note
however, that the number of stars with low or negative $v_{\phi}$
drops quickly at [Fe/H]$>-0.2$. At large negative velocities,
i.e. $v_{\phi}<-100$ km s$^{-1}$, an over-density with
$-2<$[Fe/H]$<-1.3$ is visible, corresponding most likely to the debris
from the so-called Sequoia event \citep[see][]{Sequoia}. Finally, the
fourth and the last panel in the top row gives the schematic
representation of the main features described above. The boundaries of
the ``thick'' and ``thin'' disc sequences wrap tightly around the
chevron $>$-sign feature seen in the previous panel, while the
ellipses marking the locations of the two main ancient accretion event
are more approximate. Finally, a vertical rectangular box marks the
boundaries of a section of the $v_{\phi}$, [Fe/H] space designated as
``Splash''. The Splash stars have $-0.7<$[Fe/H]$<-0.2$. This box's
metal-rich boundary corresponds to the metallicity where the number of
stars with retrograde velocities quickly drops (see first and third
panels in the top row). The rationale for choosing the metal-poor
boundary $-0.7<$[Fe/H] is laid out below.

The middle row of Figure~\ref{fig:grid} presents the orbital
properties of the stars in our sample. The disc and the accreted halo
populations have demonstrably distinct colors in most panels of this
row. For example the disc stars have large azimuthal velocities and
low radial (first panel) and vertical (second panel) dispersions. The
majority of stars with low azimuthal velocities $v_{\phi}<100$ km
s$^{-1}$ and low metallicity ([Fe/H]$<-1$) possess much higher
velocity dispersions and clearly belong to the accreted halo. Note
however, that moving from low [Fe/H] towards higher [Fe/H] at constant
$v_{\phi}\sim0$ km s$^{-1}$, the bulk kinematics change dramatically
around [Fe/H]$\sim-0.7$. For example, as shown in the first panel of
the middle row of the figure, the radial velocity dispersion drops
here from $~\sim180$ km s$^{-1}$ to $\sim110$ km s$^{-1}$. Similarly,
the vertical velocity dispersion drops from $\sim80$ km s$^{-1}$ to
$\sim60$ km s$^{-1}$; these changes imply a smaller radial and
vertical extent. This clear switch in the kinematics at
[Fe/H]$\sim=-0.7$ motivates our choice of the low-metallicity boundary
for the population we have designated the ``Splash''.

Note that this boundary demarcates not only stars with different
kinematics, but also distinct orbital properties. As seen in the
fourth panel in the middle row, the orbits of metal-poor stars with
low angular momentum can reach Galactocentric radii of $\sim15$
kpc. Stars with higher [Fe/H], on the other hand, tend to turn around
at $\sim9$ kpc, in agreement with the radial velocity behavior
described above. The fact that the Splash stars are more centrally
concentrated is also supported by the distribution of the
instantaneous radial positions shown in the third panel of the middle
row. Here, the typical value for the GS stars is $\sim8$ kpc, i.e. the
Solar radius and the centre of our selection box, implies that their
radial scale length is much larger than the extent of our sample. The
metal-rich Splash population on the other hand, is typically found
inside the Solar radius with $R\sim7$ kpc.

\begin{figure*}
  \centering
  \includegraphics[width=0.99\textwidth]{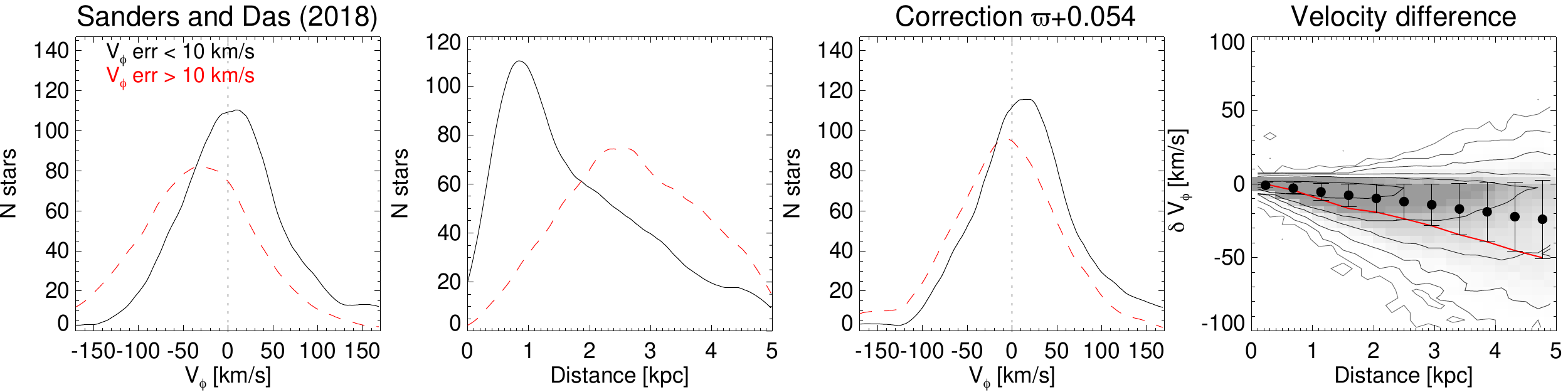}
  \caption[]{Mean azimuthal velocity of stars with large radial
    motion. Apart from the conditions described in
    Section~\ref{sec:sample} only stars with $-2<$[Fe/H]$<-0.7$,
    $0<|z|$ (kpc)$<3$ and, most importantly, $|v_{\rm R}|>250$ km
    s$^{-1}$ are shown. Black (red) solid (dashed) lines show the
    distributions for stars with low (high) $v_{\phi}$
    uncertainty. Note that for the latter we remove the $v_{\phi}$
    error $<20$ km s$^{-1}$ constraint. {\it First panel:} $v_{\phi}$
    distributions obtained with Kernel Density Estimation (with
    optimal smoothing parameter) employing Epanechnikov kernel
    \citep[see e.g.][]{Koposov2011}. Note the strong disagreement
    between the low-uncertainty (small positive ${\bar v_{\phi}}$) and
    high-uncertainty (clearly negative ${\bar v_{\phi}}$)
    samples. {\it Second panel:} Heliocentric distance distributions
    for the low- and high-uncertainty samples. The high-uncertainty
    sample is dominated by stars with large distances. {\it Third
      panel:} $v_{\phi}$ distributions after applying the
    $\varpi+0.054$ mas correction advocated by
    \citet{Schonrich2019}. Note that most of the retrograde motion
    seen in the first panel has vanished and the two distributions
    show much better agreement. {\it Fourth panel:} Azimuthal velocity
    difference $v_{\phi}-v^{\varpi+0.054}_{\phi}$ as a function if
    distance. Over-estimating stellar distances leads to a pronounced
    retrograde bias, which rapidly increases with
    distance. Data-points with error bars show the mean bias and its
    standard deviation in the given distance bin. Red curve gives the
    mean $v_{\phi}$ bias for stars with $v_{\phi}$ error $>10$ km
    s$^{-1}$.}
   \label{fig:srot}
\end{figure*}

The bottom row of Figure~\ref{fig:grid} strengthens the view that the
Splash stars are subtly different from both the Galactic halo in
general and the GS, as well as the Galactic ``thick'' and ``thin''
discs. As indicated by the first panel, while the orbital
eccentricities of the bulk of the disc stars is $e<0.3$, the
metal-rich low-angular-momentum stars have $e>0.5$, with many objects
at $e\sim1$. In the Solar neighborhood, probed by the
\citet{SandersDas2018} sample, the Splash stars have typically lower
$z_{\rm max}\sim4$ kpc (the peak Galactic height), compared to the
stars with [Fe/H]$<-0.7$ that on average can reach $z_{\rm max}\sim8$
kpc. There exists a clear distinction in terms of detailed chemical
abundances (third panel) and ages (fourth panel). The majority of the
metal-poor halo stars have high $[\alpha/\mathrm{Fe}]$ abundance
ratios, typically with $[\alpha/\mathrm{Fe}]>0.3$. Only at
-0.9$<$[Fe/H]$<$-0.7, the median $[\alpha/\mathrm{Fe}]$ of the GS
starts to decrease, diminishing slightly to $0.2-0.25$ in agreement
with \cite{Helmi2018} and \cite{Mackereth2019}. It is not surprising
that compared to the typical accreted halo stars, the median
$[\alpha/\mathrm{Fe}]$ of the Splash stars is lower at
$[\alpha/\mathrm{Fe}]\sim0.25$ given their higher metallicity. It is
nonetheless instructive to compare the $[\alpha/\mathrm{Fe}]$ of the
Splash stars to those of the disc. At fixed metallicity, the Splash
stars have higher $[\alpha/\mathrm{Fe}]$, indicating the prevalence of
the older stars in this population. The switch in
$[\alpha/\mathrm{Fe}]$ happens around $v_{\phi}\sim100$ km s$^{-1}$,
i.e. at the lower boundary of the thick disc box and the upper
boundary of the Splash box. The fourth and final panel in the bottom
row of Figure~\ref{fig:grid} shows the median age distribution. The
stellar halo hosts the oldest stars in the Galaxy with age $>12$ Gyr.
This can be compared to the thin disc ages that are predominantly
lower than $\sim6$ Gyr. The thick disc shows a strong age gradient as
a function of $v_{\phi}$ in agreement with the picture presented in
\citet{Schonrich2017}, \citet{Kawata2018} and \citet{Minchev2019}. The
metal-rich low-angular-momentum Splash population is slightly younger
than the bulk of the stellar halo and is typically as old as the
oldest stars in the thick disc.

\section{The halo and the Splash}
\label{sec:sauspl}

\subsection{Velocity ellipsoids}

The velocity ellipsoids of the halo-like stars in the Gaia Sausage and
the Splash components can be measured using a Gaussian mixture model of the entire
velocity distribution
\citep[see][for details]{Sausage}. Figure~\ref{fig:vel_fit} presents the
projections of the observed velocity ellipsoids onto the $v_{\phi}-v_{\rm
  R}$ plane together with our three-component models. 
We give the best-fit model parameters in Table~\ref{tab:comp}. The difference
between the velocity ellipsoids of the Splash (top) and the Sausage
(bottom) are evident already in the (first) left column of the
Figure. The second column shows the 2D distribution of the model
residuals (reported in numbers of Poisson sigmas). Curiously, even for
the very metal-rich disc-dominated sample shown in the top row (stars
with $2<z<3$ and $-0.7<$[Fe/H]$-0.2$), the 3-component Gaussian model does a
reasonable job, in particular around $v_{\phi}\sim0$ km s$^{-1}$ as can
also be seen in the fourth (right) column of the Figure. 1D $v_{\phi}$
projections of the data and the model are shown in the third
column. The `''thin'' (``thick'') disc components are shown with
dashed-dotted (dashed) lines. There is a pronounced ``knee'' in the
observed $v_{\phi}$ distribution around $v_{\phi}\sim 0$ km s$^{-1}$
indicating a need for an additional component. The additional Splash
component is shown with a light-blue solid line. When fitting, we fix the mean $v_{\phi}$ of the Splash component at
different trial values and choose the solution which delivers
well-behaved residuals for $v_{\phi}<100$ km s$^{-1}$ (see the top right panel of Figure~\ref{fig:vel_fit}). This procedure yields a
mean $v_{\phi}=25$ km s$^{-1}$ for the Splash. 

\begin{table}
\caption{Properties of the Gaussian mixture components}
\begin{center}
\begin{tabular}{cccccc}
Component  & fraction & $<v_{\phi}>$ & $\sigma_{\rm R}$ & $\sigma_{\phi}$  & $\sigma_{\rm Z}$\\
 & \% & km s$^{-1}$ & km s$^{-1}$ & km s$^{-1}$ & km s$^{-1}$\\
\hline
\multicolumn{6}{|c|}{$2<|z|$ (kpc)$<3$ and $-0.7<$[Fe/H]$<-0.2$} \\
\hline
``Thin'' & 25 & 220 & 31$\pm$6 & 21$\pm$4 & 26$\pm$5 \\
``Thick'' & 68 & 165 & 73$\pm$7 & 47$\pm$6 & 48$\pm$6 \\
``Splash'' & 7 & 25 & 108$\pm$19 & 54$\pm$11 & 79$\pm$21 \\
\hline
\multicolumn{6}{|c|}{$1<|z|$ (kpc)$<3$ and $-1.5<$[Fe/H]$<-1$} \\
\hline
``Thick'' & 42.5 & 160 & 73$\pm$14 & 54$\pm$9 & 64$\pm$14 \\
``Sausage'' & 50 & 0 & 175$\pm$26 & 68$\pm$15 & 82$\pm$13 \\
``Halo'' & 7.5 & 0 & 190$\pm$50 & 164$\pm$50 & 160$\pm$50 \\
\hline
\end{tabular}
\label{tab:comp}
\end{center}
\end{table}
\begin{figure*}
  \centering
  \includegraphics[width=0.98\textwidth]{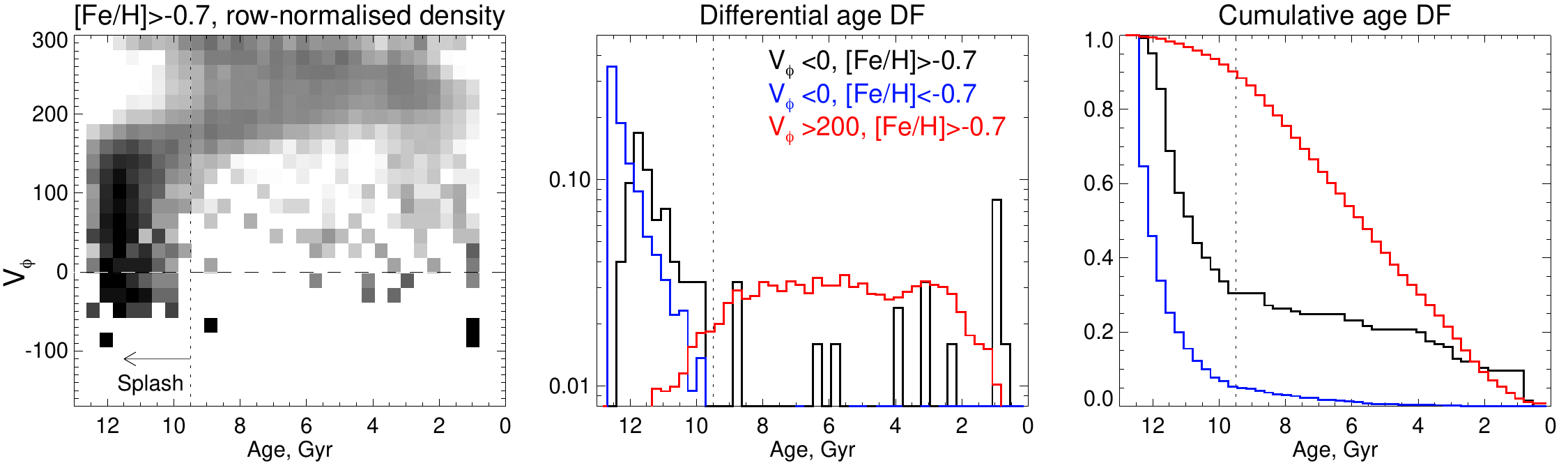}
  \caption[]{{\it Left:} Azimuthal velocity of the metal-rich stars
    $-0.7<$[Fe/H]$<-0.2$ as a function of their age. Note that the
    retrograde stars are predominantly found with ages between 12.5
    and 9.5 Gyr. These $v_{\phi}<0$ km s$^{-1}$stars are found at the
    base of a vertical over-density of stars corresponding to the
    Splash. This low angular momentum population connects seamlessly
    to the disc at $v_{\phi}>100$ km s$^{-1}$. {\it Middle:} Age
    distributions of the Sausage (blue), the Splash (black) and the
    disc (red). The major merger debris pile-up against the age grid
    boundary at 12.5, while the heated disc stars peak 1 Gyr
    later. Although the shapes of the age distributions are different,
    both are truncated at around 9.5 Gyr, which we posit marks the end
    of the Sausage merger. {\it Right:} The cumulative age
    distributions. Note a clear inflection at 9.5 in both the black (Splash)
    and blue (GS) lines.}
   \label{fig:age}
\end{figure*}

The bottom row of Figure~\ref{fig:vel_fit} gives the results of the
velocity distribution modelling for the stars with $1<z<3$ and
$-1.5<$[Fe/H]$<-1$. An even stronger ``knee'' can be seen in the third
panel, corresponding to the GS stars (light-blue solid line in the
third panel). The residual distributions shown in the second and
fourth panels are much more well-behaved since our Gaussian mixture
does not need to reproduce the complicated (and non-Gaussian) disc
kinematics. Between the top and the bottom row, there is one Gaussian
component in common, namely that corresponding to the ``thick''
disc. Re-assuringly, we recover virtually identical ``thick'' disc
Gaussian parameters as indicated by the blue dashed line shown in the
third column of the Figure. The third (very low-amplitude) Gaussian
here is the nearly-isotropic ``other halo'' component shown with
purple dashed-dotted line.

From Table~\ref{tab:comp}, we see that, for $2<z<3$ and
$-0.7<$[Fe/H]$<-0.2$, the Splash only contributes $7\%$ of all
stars. This can be compared to the GS component, which provides $50\%$
of all stars with $1<z<3$ and $-1.5<$[Fe/H]$<-1$. However, taking into
account the total number of objects in either sub-sample, the Splash
population is almost as numerous as the Sausage (notwithstanding the
differences in the Galactic height convolved with the catalogue
selection effects). In terms of the radial and vertical velocity
dispersions, the Splash is hotter by some $\sim30$ km s$^{-1}$
compared to the thick disc, but has an identical (within the
uncertainties) azimuthal velocity dispersion of $\sigma_{\phi}\sim50$
km s$^{-1}$. Compared to the GS stars, the radial velocity dispersion
is some $\sim75$ km s$^{-1}$ smaller but the vertical velocity
dispersion is similar at $\sigma_{\rm Z}\sim80$ km s$^{-1}$.

\subsection{The angular momentum of the last major merger}

The outcome of the collision between the progenitor of the {\it Gaia}
Sausage and the proto-Galaxy should depend sensitively on the geometry
of the interaction. While, most probably, the dwarf's orbit evolved
rapidly under the effects of dynamical friction, we can glean the
final state of the impactor's angular momentum by measuring the spin
of the dominant stellar halo component. The multi-Gaussian velocity
field decomposition described above yields $v_{\phi}\sim0$ km s$^{-1}$
for the Sausage component, implying no appreciable angular momentum
for the last major debris as measured in the dataset of
\citet{SandersDas2018}. This can be compared with the small but
statistically significant rotation of $\sim20-30$ km s$^{-1}$ reported
by \citet{Sausage} using a dataset with a similar reach, i.e. a few
kpc from the Sun. This estimate based on the SDSS-Gaia-DR1 dataset was
recently confirmed by \citet{Tian2019} using nearby LAMOST K-giants in
the {\it Gaia} DR2. On much larger scale, namely tens of kpc
throughout the Galaxy, the net stellar halo rotation was measured by
\citet{SlightSpin} who used three different tracers samples, namely RR
Lyrae, Blue Horizontal Branch stars and K giants to arrive at a
$v_{\phi}\sim15$ km s$^{-1}$. This is in agreement with the more
recent study of the RR Lyrae kinematics carried out by
\citet{Wegg2019}.

The above studies are clearly at odds with the claim of
\citet{Helmi2018} of a strongly-retrograde motion of the bulk of the
stellar halo locally. Note that this is not the first detection of a
retrograde stellar halo motion: for instance, \citet{Carollo2007}
measure a small prograde rotation for the ``inner halo'' but report a
net retrograde motion for the ``outer halo'' component (also see
references to the earlier studies in their work). However according to
\citet{AllegedDuality}, this apparent retrograde spin may largely be
due to over-estimated distances. Note that it is now established with
certainty that the {\it Gaia} DR2 parallaxes are biased low, implying that
the distances are over-estimated. For example, \citet{Lindegren2018}
report a bias of $-0.029$ mas from quasars, but warned that the offset could be dependent on on-sky position, magnitude and colour. \citet{Schonrich2019} measure a
stronger bias of $-0.054$ mas with a small uncertainty \citep[in
  agreement with][]{Zinn2019} for the Gaia RVS sample. Moreover they discover the dependence
of the parallax offset on the parallax error, thus exacerbating the
bias for distant stars. \cite{SandersDas2018} did not apply an offset before deriving output parallaxes from the combination of astrometry and spectro-photometry. As we will see, when the spectro-photometry is highly informative the bias from the astrometry will have a weak effect, but for more uncertain spectro-photometry the output distances (and hence velocities) will inherit the Gaia biases. Due to this complex interaction between the different data sources, it is impossible to accurately correct \emph{a posteriori} for a given parallax offset.

Given the high radial anisotropy of the Sausage stellar debris (see
bottom left panel of Figure~\ref{fig:vel_fit}), it is possible to
estimate the bulk angular momentum of this halo component simply by
selecting stars with high $|v_{\rm R}|$. Accordingly,
Figure~\ref{fig:srot} shows the distribution of azimuthal velocities
for the stars in our main sample (see above) but limited to
$-2<$[Fe/H]$<-0.7$, $0<|$z$|$ (kpc)$<3$ and, most importantly,
$|v_{\rm R}|>250$ km s$^{-1}$. Additionally, we split this sample into
two parts of approximately the same size ($\sim1200$ stars in each):
one with small $v_{\phi}$ errors $<10$ km s$^{-1}$ (black solid lines)
and one with $v_{\phi}$ errors in excess of $10$ km s$^{-1}$ (red
dashed lines). As evidenced by the Figure, the mean spin of the two
samples is unmistakably different: the peak of the low-uncertainty
sample is at around $+10$ km s$^{-1}$ while the high-uncertainty
distribution peaks at approximately $-40$ km s$^{-1}$. The cause of
such a drastic difference is hinted at in the second panel of
Figure~\ref{fig:srot} which shows the heliocentric distance
distributions of the two samples. The low-uncertainty sample is
largely limited to the nearby stars with distances of order of 1-2
kpc. The high-uncertainty sample is composed of stars with a median
distance of $\sim3$ kpc. Note that in accordance with our base
selection, all of these stars have $|\varpi/\sigma_{\varpi}|>5$ (see
Section~\ref{sec:sample}).

\begin{figure*}
  \centering
  \includegraphics[width=0.99\textwidth]{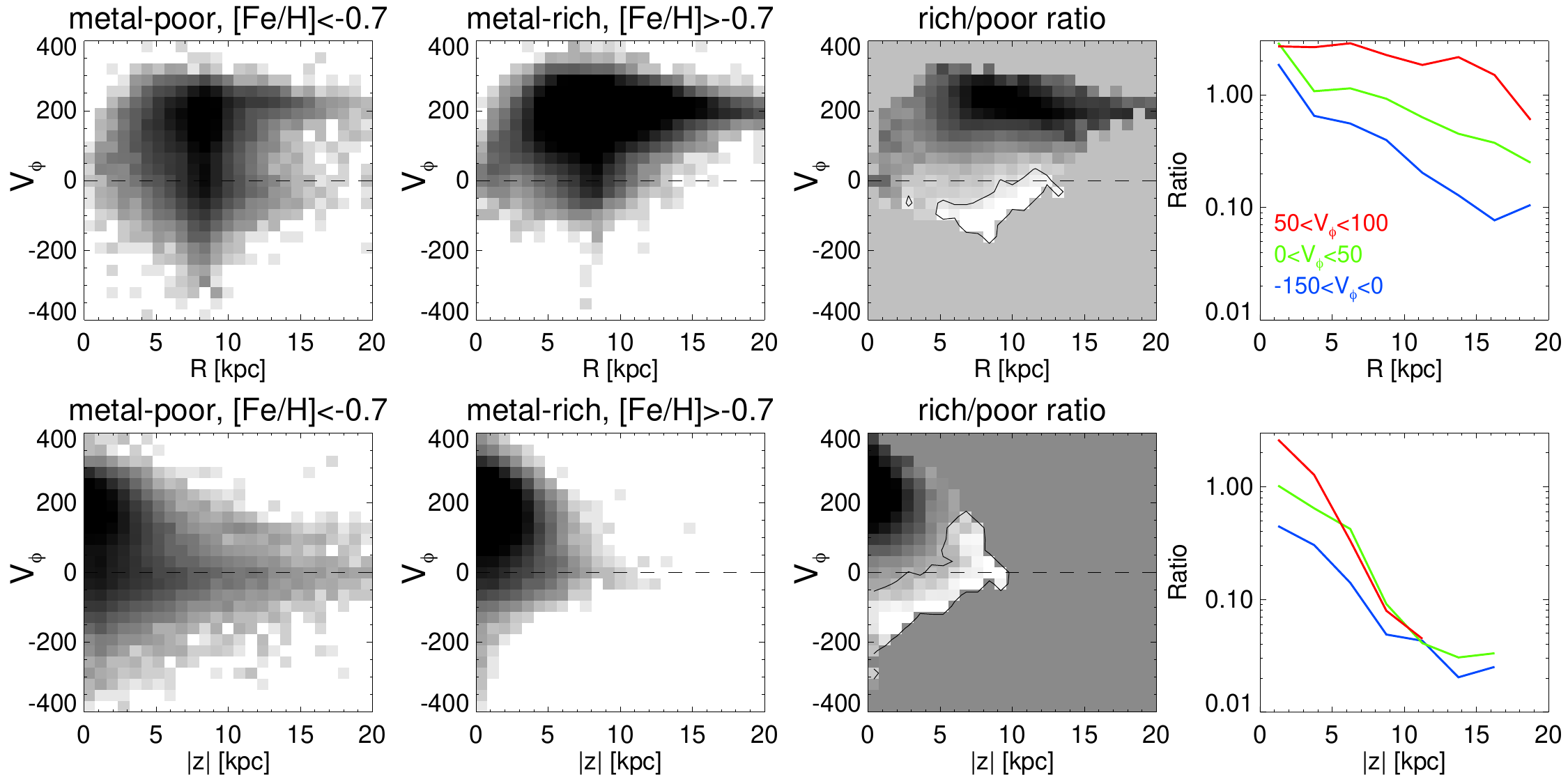}
  \caption[]{Metal-poor and metal-rich stars in the plane of
    $v_{\phi}$ vs $R$ (top) and $|z|$ (bottom). Note that while our
    sample is dominated by nearby stars, a small number of objects
    spans a wide range of $R$ and $|z|$. The selection effects are
    strong and increase with distance. However we hypothesise that
    they affect in equal measure the metal-poor (first column) and the
    metal-rich (second column) sample. Thus their ratio (third column)
    may show the true intrinsic variation of the density ratio as a
    function of $R$ and $|z|$. Note that the overlaid contour in the
    third panel corresponds to $50\%$ ratio of metal-rich to
    metal-poor star counts. The fourth column displays the evolution
    of the density ratio for stars in three $v_{\phi}$ bins, starting
    with purely retrograde (blue), through low net rotation (green) to
    prograde (red). As clear from both the third and the fourth
    column, the Splash population decays much faster with Galactic
    height and at $|z|\sim10$ kpc drops to a meager $\sim5\%$ of the
    halo density.}
   \label{fig:ratio}
\end{figure*}

\citet{Schonrich2019} use the procedures described in
\citet{Schonrich2012} and \citet{SchonrichAumer2017} to create a large
sample of unbiased distance estimates based on the {\it Gaia} DR2
parallaxes for the RVS sample. Here, we only use one of the components
of their method, namely the global parallax correction of $0.054$ mas
to investigate the changes in the mean angular momentum of the {\it
  Gaia} Sausage. The results of increasing parallax by $0.054$ mas are
shown in the third panel of Figure~\ref{fig:srot}. The distribution of
the azimuthal velocities in the low-uncertainty sample shows a small
shift towards higher mean prograde motion, with its peak now at $\sim
+20$ km s$^{-1}$. The $v_{\phi}$ distribution of the high-uncertainty
sample exhibits a much more dramatic evolution on applying the
parallax correction. The distribution tightens up and the peak moves
to $\sim-10$ km s$^{-1}$, displaying overall much better agreement
with the low-uncertainty sample. As illustrated in the fourth panel of
the Figure, the principle cause of the difference in the behaviour of
the two samples is the distance-dependent $v_{\phi}$ bias caused by
using under-estimated {\it Gaia} DR2 parallaxes. According to the
distribution of $v_{\phi}-v^{\varpi+0.054}_{\phi}$, stars at 3 kpc
from the Sun typically display $-15$ km s$^{-1}$ spurious retrograde
motion. Note however, that for stars with large $v_{\phi}$ errors (red
line) this bias is twice as large $-30$ km s$^{-1}$ at 3 kpc, and the
tails of the distribution reach $-100$ km s$^{-1}$. The observed bias
is likely larger than that shown in the fourth panel of
Figure~\ref{fig:srot} as the parallax systematics appear to correlate
with the parallax errors \citep[see][]{Schonrich2019}.

Therefore, we conclude that the net retrograde motion of the
radially-anisotropic local sample claimed by \citet{Helmi2018} is
limited to the stars with larger azimuthal velocity uncertainties and
may be caused by over-estimated distances. We demonstrate that even
without accounting for the parallax bias, but limiting the sample to
high-accuracy $v_{\phi}$ measurements from the \cite{SandersDas2018}
catalogue, one obtains mean $v_{\phi}\sim+10$ km s$^{-1}$ for the {\it
  Gaia} Sausage debris. On including the correction, this estimate
shifts slightly to $v_{\phi}\sim+20$ km s$^{-1}$.

\subsection{Dating the last major merger with the Splash}

In the metallicity range of the
Splash, there appears to be a wide range of ages and angular momenta,
not necessarily captured adequately by the median distribution shown
in Figure~\ref{fig:grid}. Accordingly, Figure~\ref{fig:age} explores further
correlations between the angular momentum and age for stars with
$-0.7<$[Fe/H]$<-0.2$. Note that as described earlier, we apply an
additional stellar population restriction for the age study, namely
selecting only stars around the MSTO. The Figure presents
row-normalized density in the space of $v_{\phi}$ and age, which
allows us to track the typical age for a given range of azimuthal
velocity. As clear from the left panel of the Figure, for
$v_{\phi}>150$ km s$^{-1}$, a wide range of ages is plausible,
highlighting the fact that the star-formation started in the disc as
early as 12 Gyr and proceeded to the present day.

\begin{figure*}
  \centering
  \includegraphics[width=0.97\textwidth]{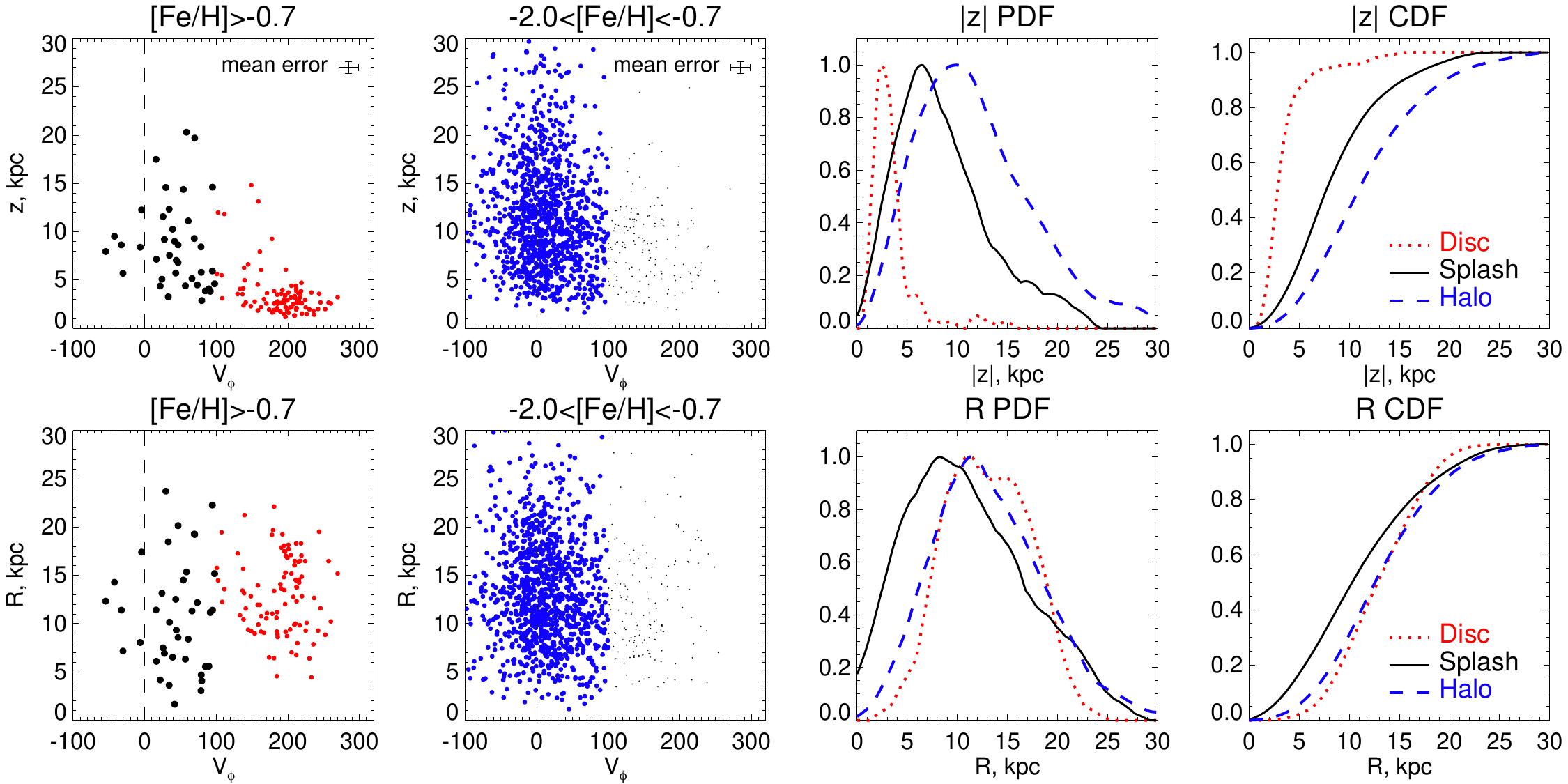}
  \caption[]{SDSS K-giants in the Splash (black, solid), disc (red,
    dotted) and halo (blue, dashed). All distributions are obtained
    using KDE with Epanechnikov kernel and optimal scale. {\it Top
      row:} $|z|$ distributions. {\it First panel:} $|z|$ vs azimuthal
    velocity $v_{\phi}$ for stars with [Fe/H]$>-0.7$. {\it Second
      panel:} Same as previous panel but for stars with
    $-2<$[Fe/H]$<-0.7$. {\it Third panel:} probability distributions
    for the three populations selected as shown in the first two
    panels. The disc extent is smaller than that of the Splash, which
    in turn is not as large as that of the halo. {\it Fourth panel:}
    Cumulative distributions. {\it Bottom panel:} $R$
    distributions. Panels are the same as above but for $R$ instead of
    $|z|$. As the third and the fourth panels demonstrate, the radial
    size of the Splash is smaller that that of the disc and the halo,
    which match each other in 2-D projected distance as traced by the
    SDSS K-giants.}
   \label{fig:kgrid}
\end{figure*}
\begin{figure}
  \centering
  \includegraphics[width=0.499\textwidth]{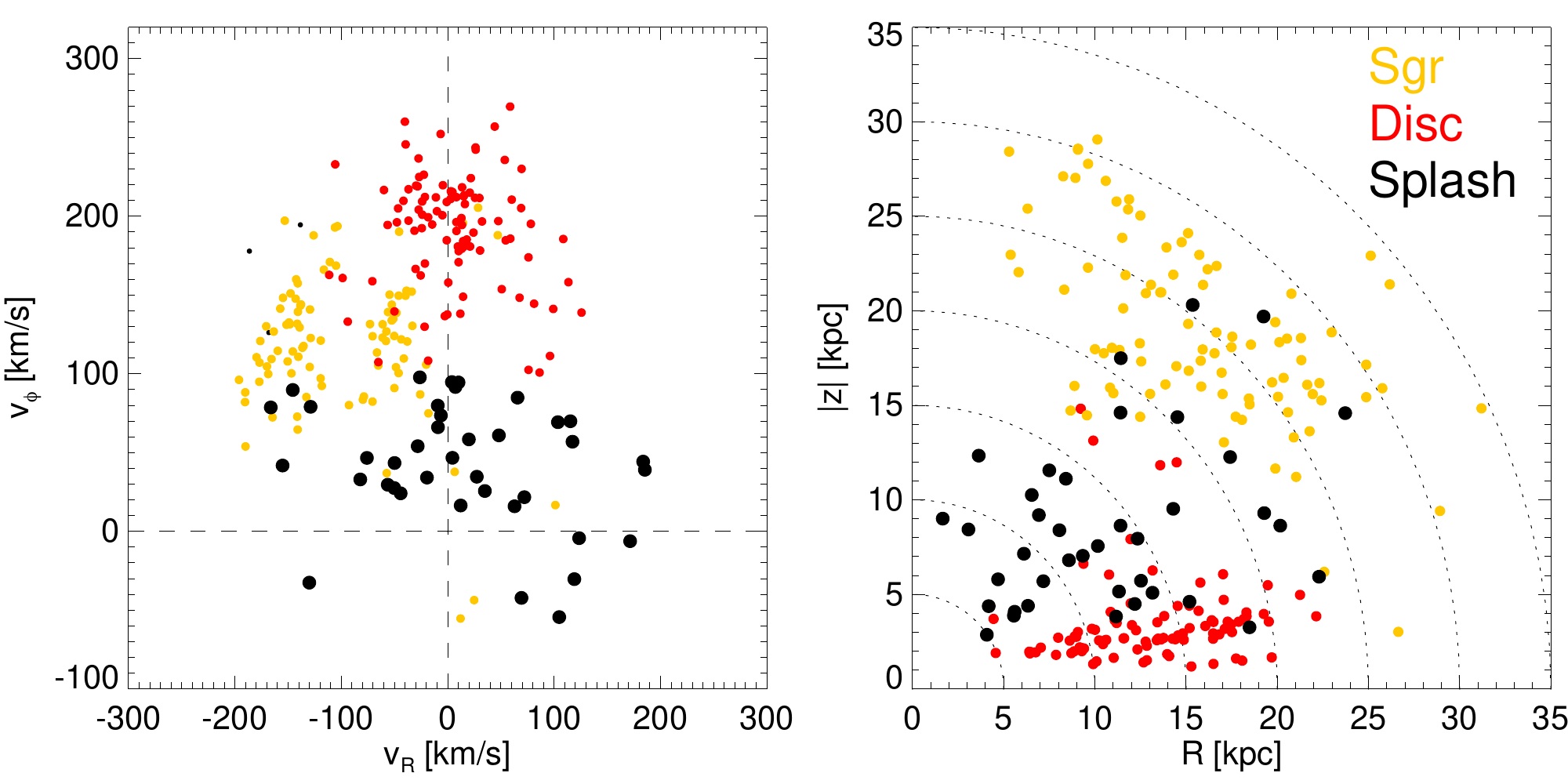}
  \caption[]{Metal-rich K-giants in the Sgr tidal stream (yellow) are
    compared to those in the Splash and the disc. Sgr K-giants are
    selected to have the same metallicity, but can clearly be seen
    extending much further in $|z|$ and/or $R$.}
   \label{fig:kgzr}
\end{figure}

The age distribution for the slower rotating stars, i.e. those with
$v_{\phi}<150$ km s $^{-1}$ is clearly very different. In particular,
the metal-rich retrograde (those with $v_{\phi}<0$ km s$^{-1}$) stars
are predominantly found with ages greater than 9.5 Gyr. In the left
panel of Figure~\ref{fig:age}, these retrograde stars are at the base
of an extended column-like feature which stretches over a wide range
of azimuthal velocities, from $v_{\phi}\sim-100$km s$^{-1}$ to
$v_{\phi}\sim+150$ km s$^{-1}$ in the relatively narrow range of ages
from 9.5 to 12.5 Gyr. This dark vertical band of stars with
$v_{\phi}<100$ km s$^{-1}$ and old ages is the Splash. From this
Figure, the close connection between the Splash and the disc is
immediately apparent: it is not clear where the disc stops and the
Splash begins. The two clearly overlap at
$100<v_{\phi}/\mathrm{km s^{-1}} <150$. Additionally, as evidenced from the Figure, Splash
stars display net prograde rotation, albeit with a lower amplitude than the disc.

Given the seamless transition from the disc to the Splash population
observed in the left panel of Figure~\ref{fig:age} as well as in the
multiple panels of Figure~\ref{fig:grid}, we hypothesize that Splash
is the population of stars originally born in the proto-disc of the
Galaxy and subsequently kicked (splashed) into low-angular-momentum
(high eccentricity) orbits by an accretion event that finished around
9.5 Gyr ago. The best candidate for such an event is the {\it Gaia}
Sausage merger. We compare the age make-up of the heated disc stars
and that of the accreted Sausage stars in the middle panel of
Figure~\ref{fig:age}. We limit the Splash stars to those with
$v_{\phi}<0$ km s$^{-1}$ to minimize any disc contamination. The
accreted stars are also selected to have $v_{\phi}<0$ km s$^{-1}$ but
are more metal-poor than [Fe/H]$=-0.7$ (see Figure~\ref{fig:grid} for
reference). The accreted stars are dominated by the oldest stars in
our sample, with the distribution function quickly dropping as a
function of formation epoch, with no clear signal discernible younger
than 9.5 Gyr. The Splash age distribution looks sufficiently different
from that of the {\it Gaia} Sausage. For example, its peak is not at
12.5 Gyr but 1 Gyr later, which is understandable as even for
something as massive as the Milky Way progenitor, it takes a
non-trivial amount of time to self-enrich to [Fe/H]$\sim-0.5$. Even
though the shapes of the age distribution of the Sausage and Splash
stars are rather different, they have one particular feature in
common: a truncation at 9.5 Gyr.

We thus use the synchronicity between the cessation of star-formation
in the {\it Gaia} Sausage and the finishing of the disc heating in the
Milky Way to put constraint on the epoch of the last major merger
event.

\section{Splash beyond the Solar neighborhood}
\label{sec:beyond}

\begin{figure*}
  \centering
  \includegraphics[width=0.98\textwidth]{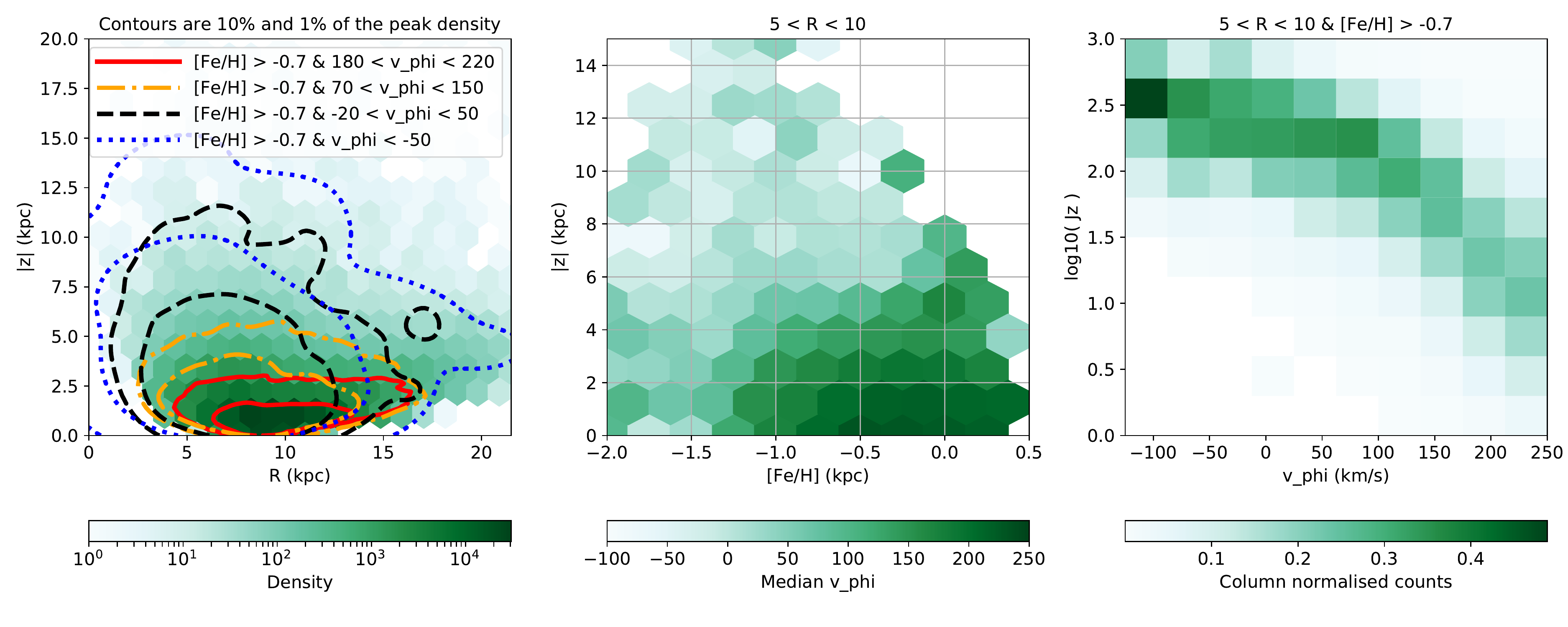}
  \caption[]{Using LAMOST K-giants to probe the spatial extent of the Splash. The left panel shows the distribution of metal-rich stars of various $v_\phi$. The middle panel shows the variation of $v_\phi$ with $|z|$ and [Fe/H], illustrating how at the intermediate metallicity ($-1 < {\rm [Fe/H]} < -0.3$) population transitions from being dominated by the higher angular momentum thick disc at $|z| \la 5$ kpc to being dominated by the lower angular momentum Splash at $|z| \ga 5$ kpc. The right panel shows that for the higher angular momentum stars there is a strong correlation between $v_\phi$ and $J_z$, but for the Splash stars ($v_\phi \la 50$ km/s) this trend appears to flatten.}
   \label{fig:kg_lamost_spatial}
\end{figure*}
\begin{figure}
  \centering
  \includegraphics[width=0.45\textwidth]{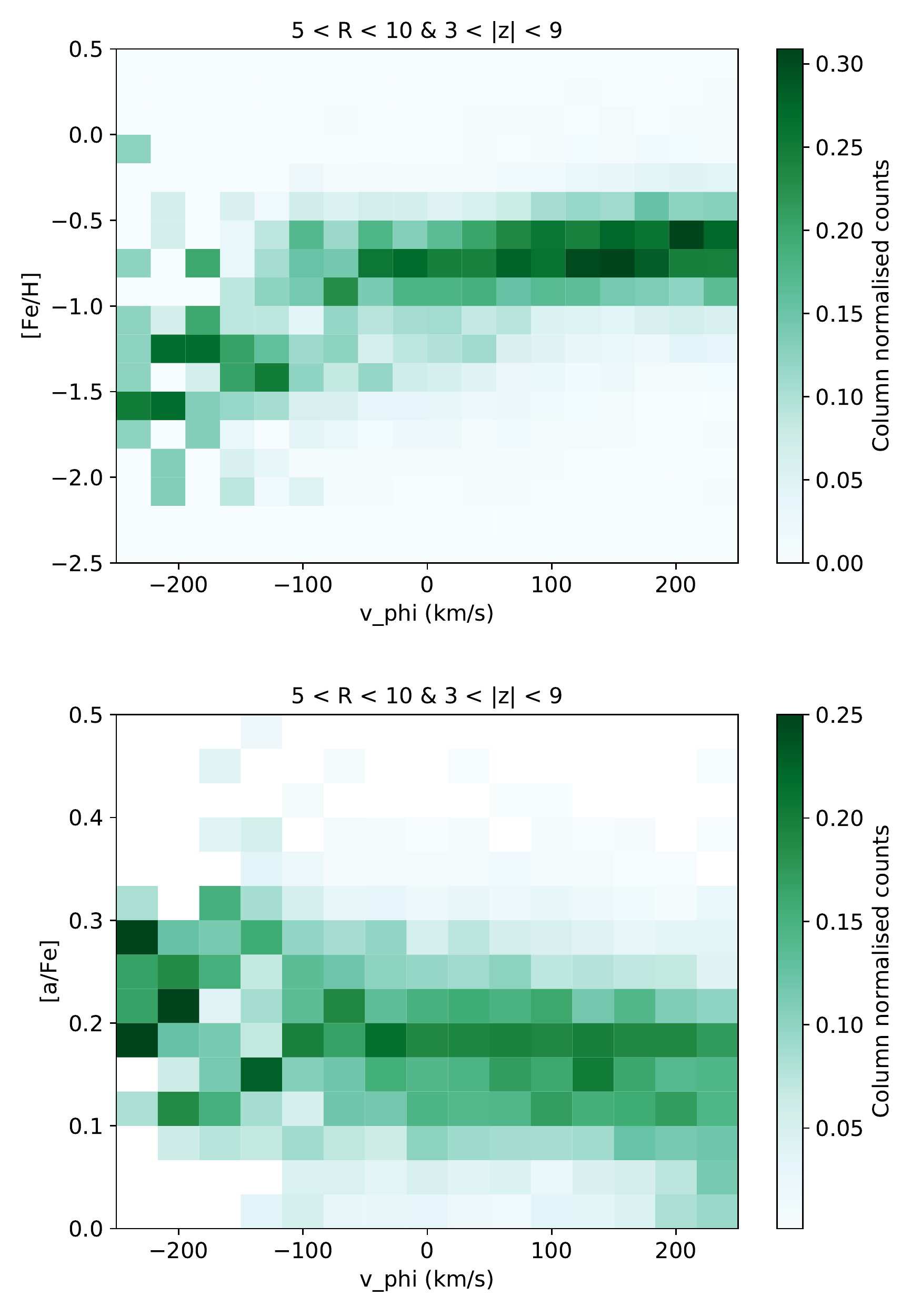}
  \caption[]{
  The chemical properties of the Splash for the LAMOST K-giants, concentrating on the interface between the thick disc and the Splash ($3 < |z| < 9$ kpc). The upper panel shows how the intermediate metallicity stars, which appear to be a chemically homogeneous population, extend to negative $v_\phi$.
  The lower-panel, which shows the alpha-element abundance of these stars, again demonstrates the  chemically homogeneous nature of stars over a wide span of $v_\phi$.}
    \label{fig:kg_lamost_chemistry}
\end{figure}

The implications of the detection of a large number of stars with
[Fe/H]$>-0.7$ on highly eccentric and relatively energetic orbits in
the Solar neighborhood are clear: there ought to be a corresponding
population of metal-rich stars at large Galacto-centric radii and
heights. In what follows we first explore the evolution of the density
ratio of the metal-poor and metal-rich low-angular-momentum stars in
our primary sample (described above) and then probe the existence of
such a population using the SDSS and LAMOST K-giant stars.

\subsection{Metal-rich to metal-poor density ratio}

The parent catalogue of \citet{SandersDas2018} comprises data from
several spectroscopic surveys, each with its own strong selection
effects. These biases are predominantly spatial, driven by each
survey's footprint and limiting magnitude. Here we do not attempt to
correct for the strong selection biases affecting our sample. Instead,
we assume that these biases impact the metal-rich and the metal-poor
stars equally. If this conjecture is correct, the ratio of the
metal-rich and metal-poor star counts may be able to reveal the
intrinsic evolution of the density of the Splash stars normlized by
that of the halo. Note that in the volume probed by our sample, the
metal-poor halo population is dominated by the {\it Gaia} Sausage
debris, which has a triaxial 3D shape \citep[see][]{Iorio2019} and a
characteristic density break around $20-30$ kpc from the centre of the
Galaxy \citep[see][]{PileUp}.

Figure~\ref{fig:ratio} shows the density of metal-poor (first column)
and metal-rich (second column) stars in the plane spanned by azimuthal
velocity $v_{\phi}$ as a function of Galacto-centric radius (top) and
Galactic height $|z|$ (bottom). As clear from the Figure, both radial
and vertical density distributions are affected by the selection
effects. However, the ratio of the density distributions shown in the
third column evolves much more smoothly revealing a particular
behavior of the Splash stars in comparison to the halo population. In
the Solar neighborhood, the ratio of the number of the metal-rich
retrograde stars to that of the metal-poor ones is close to
$1:1$. It evolves slowly as a function of Galacto-centric radius
$R$, dropping to $1:10$ at $R\sim15$ kpc. Vertically, the density
ratio evolves noticeably faster, hitting $10\%$ at $\sim7$ kpc and
$\sim5\%$ around 10-15 kpc.

If the behaviour displayed in the third and fourth columns of
Figure~\ref{fig:ratio} is not seriously affected by the selection
effects, we can conclude that the extent (both radial and vertical) of
the Splash is considerably smaller than that of the halo. Splash may
be limited to 15-20 kpc in the radial direction and 10-15 kpc above
the plane.

\subsection{The SDSS K-giants}
\label{sec:kg}

We use the sample of K-giants identified in the Sloan Digital Sky
Survey spectroscopy by \citet{Xue2014} who also provide distance
estimates to these stars. Out of $\sim$6,000 objects, only $\sim280$
have [Fe/H]$>-0.7$. Additionally, we also require the metallicity
error to be lower than 0.2 and the azimuthal velocity error to be
lower than 30 km s$^{-1}$. We take care to minimize the contamination
of our distant Splash sample by the tidal debris from the currently
disrupting Sgr dwarf. More specifically, we classify the stars with
$|B_{\rm Sgr}|<15^{\circ}$ and heliocentric distance $D>15$ kpc as
those belonging to the Sgr stream. Here, $B_{\rm Sgr}$ is the latitude
in the coordinate system aligned with the Sgr's tidal tails, namely a
great circle with a pole at (RA, Dec) $= (303.63^{\circ},
59.58^{\circ})$ \citep[see e.g.][]{Belokurov2014}. There are 94 likely
Sgr stream members with [Fe/H]$>-0.7$. Finally, to select stars with
low angular momentum we require $v_{\phi}<100$ km s$^{-1}$. After
applying the above cuts, we end up with 39 metal-rich likely Splash
K-giants at large distances from the Sun. We compare these to the disc
stars, selected with the same metallicity cut but $v_{\phi}>100$ km
s$^{-1}$ (96 stars in total) and the halo objects (likely dominated by
the {\it Gaia} Sausage debris), identified as those with
$v_{\phi}<100$ km s$^{-1}$ and $-2<$[Fe/H] $<-0.7$ (1091 stars in
total).

Figure~\ref{fig:kgrid} shows the spatial properties of the Splash
K-giants (black points and black solid lines) and those in the disc
(red points and red dotted lines) and the halo (blue points and blue
dashed lines). As the top row of the Figure demonstrates, the disc
stars are limited to $|z|<5$ kpc, while the Splash population extends
as far as $|z|\sim20$ kpc. This is not surprising given the findings
of Section~\ref{sec:sample} and \ref{sec:sauspl}. Similarly, in
agreement with the local kinematics, the halo populations extend
beyond the reach of the Splash stars, i.e. to $|z|>30$ kpc. In terms
of the radial reach, displayed in the bottom row of the Figure, the
Splash stars can travel as far as $R\sim 25$ kpc, but their $R$
distribution peaks at smaller distances from the Galactic centre
compared to both the disc and the halo. Curiously, the radial distance
distributions of the disc and the halo are nearly identical as probed
by this K-giant sample.

We emphasize that, notwithstanding the low numbers of tracers, the
ranking of the vertical and the radial sizes of the Splash, the disc
and the halo, i.e. $z_{\rm disc}<z_{\rm Splash}<z_{\rm halo}$ and
$R_{\rm Splash}<R_{\rm halo}$, is likely robust to the selection
biases of the SDSS K-giant sample. Supporting evidence can be found in
Figure~\ref{fig:kgzr} which compares the Galacto-centric positions of
the Splash and disc populations to those of the Sgr stream members
(yellow points). Sgr stars are chosen using the same metallicity cut
as the Splash and the disc, i.e. [Fe/H]$>-0.7$. As the Figure clearly
demonstrates, the SDSS K-giant sample contains plenty of metal-rich
and distant stars, i.e those with $r>25$ kpc (all in the Sgr
stream). Yet, the Splash giants are mostly limited to $r<20$ kpc, with
more than a half of our sample at $r<15$ kpc.

Overall, the behaviour of the SDSS K-giants support conclusions of the
previous sub-section. The Splash stars amount to only $3-5\%$ of the
halo density at these large Galactic radii and heights. The extent of
the Splash is much smaller than that of the halo: its reach is curbed
around 15-20 kpc in $R$ and 10-15 kpc in $|z|$. The peak of the
Splash's $R$ ($z$) distribution function is at 10 (7) kpc.

\subsection{The LAMOST K-giants}
\label{sec:kg_lamost}

In order to increase our sample size, we now explore a larger catalogue of K-giants from the LAMOST survey \citep{Luo2015}. This is constructed using stellar parameters from the SP\_Ace pipeline \citep{Boeche2018}, which has been updated using the 5th LAMOST data release.
We use stars where SP\_Ace has converged without any errors, the SP\_Ace signal-to-noise is greater than 20, $|{\rm b}|>10$ deg and the following error cuts:
${\rm \delta log(g) < 0.5}$,
${\rm \delta T_{eff} < 150\:K}$,
${\rm \delta [Fe/H] < 0.15\:dex}$,
${\rm \delta v_R < 30\:km/s}$,
${\rm \delta \mu_{\alpha} < 0.5\:mas/yr}$ and
${\rm \delta \mu_{\delta} < 0.5\:mas/yr}$ are satisfied. 
We also reject stars observed prior to {\rm MJD = 55945} as these have been shown to be problematic (see \citealt{Boeche2018}). Duplicates have been removed, retaining only the spectrum with highest signal-to-noise.
Our K-giants are selected using a cut of ${\rm log(g)} < 3.4$ and ${\rm T_{eff} < 5600\:K}$. 
To ensure we have minimal contamination from late-type dwarfs we have imposed a further cut of $M_G < 3$, where $M_G$ is calculated from the parallax and for all stars we require parallax error to be less than 0.2 arcsec. Although this parallax error cut likely removes some low signal-to-noise bona fide giants, we feel that this is warranted so as to give us as clean a sample of giants as possible.

For the stars with parallax error less than 20\% the distance has been calculated by inverting the parallax. For those with parallax error greater than 20\% distances have been calculated using a random forest regression tool trained on a high-quality subset of the data with parallax errors less than 15\%
${\rm \delta log(g) < 0.2}$,
${\rm \delta T_{eff} < 100\:K}$,
${\rm \delta [Fe/H] < 0.1\:dex}$,
${\rm |b| > 30\:deg}$ and
${\rm |z| > 0.5\:kpc}$. The penultimate cut minimises the importance of the extinction correction and the final cut is used to avoid the regression being dominated by thin disc stars. For most of these stars the distance accuracy is around 15\%, although for the brightest stars ($M_G < -2$) the errors increase and so we have excluded these from our analysis.
For all LAMOST giants the extinction correction is derived from the maps of \citet{Schlegel1998} with the additional latitude-dependent correction of \citet{RAVEDR5}. Finally, we have removed stars which are likely members of the Sagittarius stream by rejecting stars that satisfy the following three criteria: within 20 deg of the stream centre according to \citet{Belokurov2014}; within 20 km/s of the stream radial velocity at that location according to \citet{Belokurov2014}; and distance within 40\% of the distance at that location according to \citet{Hernitschek2017}. Our resulting sample consists of 256,000 stars to distances up to 20 kpc. If we focus on metal-rich stars with ${\rm [Fe/H]} > -0.7$ dex, we have 6106 with $|z| > 3 kpc$ and 779 with $|z| > 5 kpc$.

\begin{figure*}
  \centering
  \includegraphics[width=0.99\textwidth]{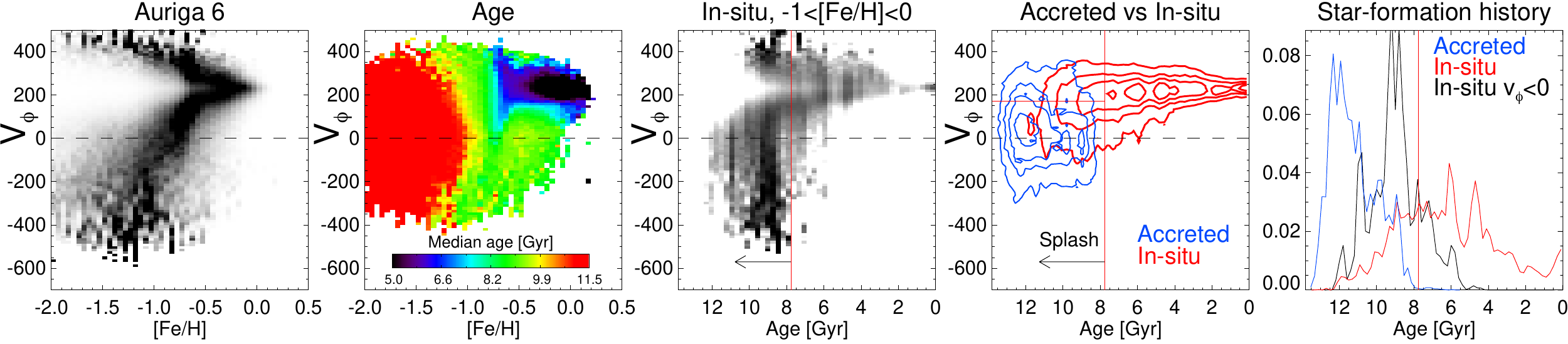}
  \includegraphics[width=0.99\textwidth]{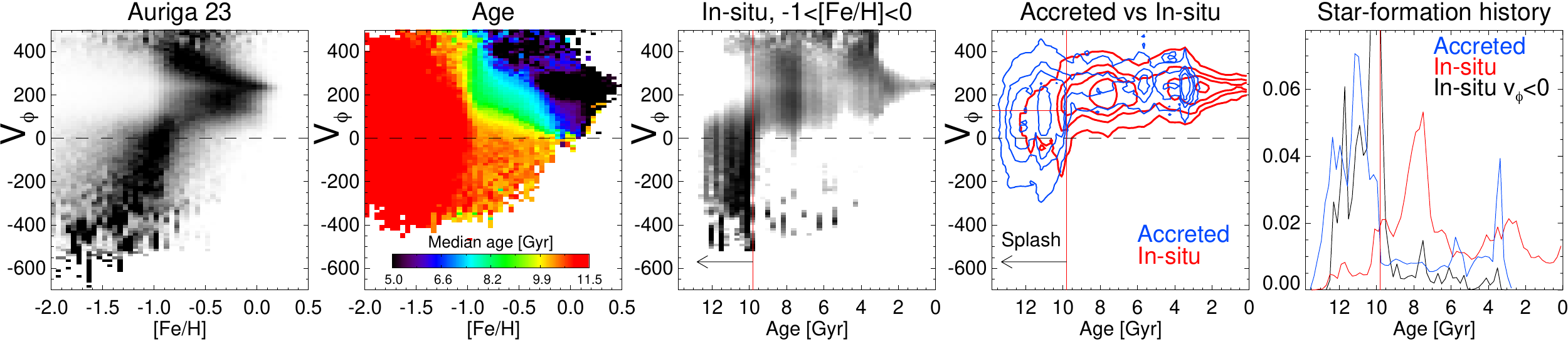}
  \includegraphics[width=0.99\textwidth]{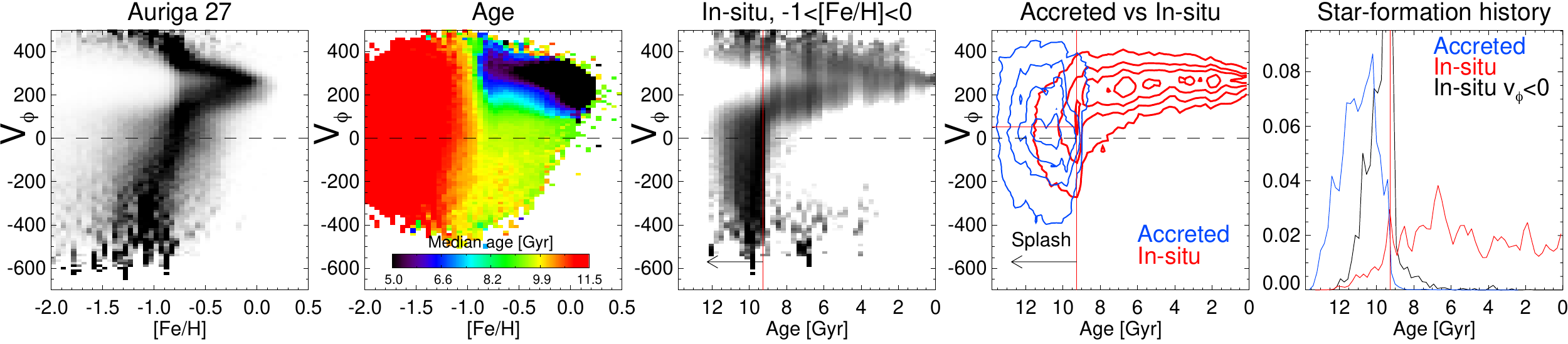}
  \caption[]{Solar neighborhood in Auriga 6 (top), 23 (middle) and 27
    (bottom). Only stars with $5<R<11$ and $0.5<|z|<3$ are
    included. {\it First column} is the numerical counterpart of top
    third panel of Figure~\ref{fig:grid} and shows row-normalized
    density in the space of $v_{\phi}$ and metallicity. Note the
    thin-thick disc chevron feature at high [Fe/H] and the low angular
    momentum overdensity at lower [Fe/H]. {\it Second column} gives
    the median age distribution in the same space (compare to the
    bottom right panel of Figure~\ref{fig:grid}). The top (thin disc)
    part of the chevron is predominantly young age. In Auriga 23 and
    27, where the last major merger happened some 10 Gyr ago, a clear
    Splash-like feature is visible for stars with [Fe/H]$>$-1 and
    $v_{\phi}<0$. {\it Third column} is the analog of the left panel
    in Figure~\ref{fig:age} and shows the dependence of $v_{\phi}$ on
    stellar age for all metal-rich stars. The vertical Splash column
    is clearly visible in all three Auriga galaxies considered
    here. Vertical red line gives an approximate young edge of the
    Splash, signifying the end of the major merger. This value is
    calculated as the 20th percentile of the age distribution of the
    metal-rich stars with $v_{\phi}<0$. Note that in all examples, the
    ``thick'' (slow) disc population continues to form after the major
    merger. {\it Fourth column} The $v_{\phi}$-age distribution is
    split into accreted (blue) and in-situ (red) components. Note that
    the edge of Splash (shown with the vertical red line, see previous
    panel) coincides perfectly with the end of the major accretion
    event. Horizontal red line gives the median $v_{\phi}$ of the
    Splash stars (i.e. those to the left of the vertical red line). In
    the case of Auriga 6 and 23, it is clear that the Galaxy was in a
    proto-disc state before the accretion as it managed to retain much
    of the angular momentum. The case of Auriga 27 is less clear, it
    could have started in spheroid (rather than a disc)
    configuration. Finally, {\it fifth column} presents the age
    distributions for the accreted (blue), in-situ (red) and Splash
    (black) stars (compare to the middle panel of
    Figure~\ref{fig:age}). In all cases, the accreted population is
    typically older than that of the Splash. In terms of the local
    star-formation activity, the SF rate typically increases during
    the merger and subsides immediately after.}
   \label{fig:auexample}
\end{figure*}

The properties of these LAMOST K-giants are illustrated in
Fig. \ref{fig:kg_lamost_spatial}. The left panel shows the spatial
distribution of all stars (greyscale), together with various subsets
of metal-rich stars. The higher angular momentum populations, with
[Fe/H]$>-0.7$ and $70 < v_{\phi} < 220 {\rm km/s}$ are clearly
confined to the plane, with $|z| \la 5$ kpc. As has been shown for the
SDSS K-giants, the Splash stars reach much higher, to around 10 kpc.
Note that since we are restricting ourselves to a specific range of
[Fe/H] we are minimising the influence of any metallicity-dependent
bias in the distances. The middle panel shows that there is a
transition in $v_\phi$ at $|z| \sim 5$ kpc, with the higher angular
momentum stars confined to smaller $|z|$. The right panel shows how
the vertical extent, as quantified by the vertical action $J_z$,
depends on $v_\phi$. Actions have been calculated using the Galpy
package \citep{Bovy2015}, adopting the MWPotential2014 model for the
potential. For the higher angular momentum stars (i.e. the `canonical'
thick disc with $v_\phi \ga 100$ km/s) there is a strong correlation
between $v_\phi$ and $J_z$, but for the Splash stars ($v_\phi \la 50$
km/s) this trend appears to flatten.

In Fig. \ref{fig:kg_lamost_chemistry} we illustrate the chemical
properties of the Splash for the LAMOST K-giants, concentrating on the
interface between the thick disc and the Splash ($3 < |z| < 9$
kpc). The upper panel shows how the intermediate metallicity stars,
which would normally be associated to the thick disc, stretch to very
low angular momenta and even extend to negative $v_\phi$. In terms of
[Fe/H] these stars appear to be a homogeneous population. A similar
conclusion can be drawn from the alpha-element abundance of these
stars, which again appears homogeneous over a wide span of
$v_\phi$. Until one reaches the halo material at $v_\phi \la -50$
km/s, most of the stars are concentrated at ${\rm [\alpha/Fe]} \sim
+0.2$, which is what one expects for the thick disc population.

\section{Discussion and Conclusions}
\label{sec:disc}

\subsection{Making a Splash}
\label{sec:sims}

\begin{figure*}
  \centering
  \includegraphics[width=0.32\textwidth]{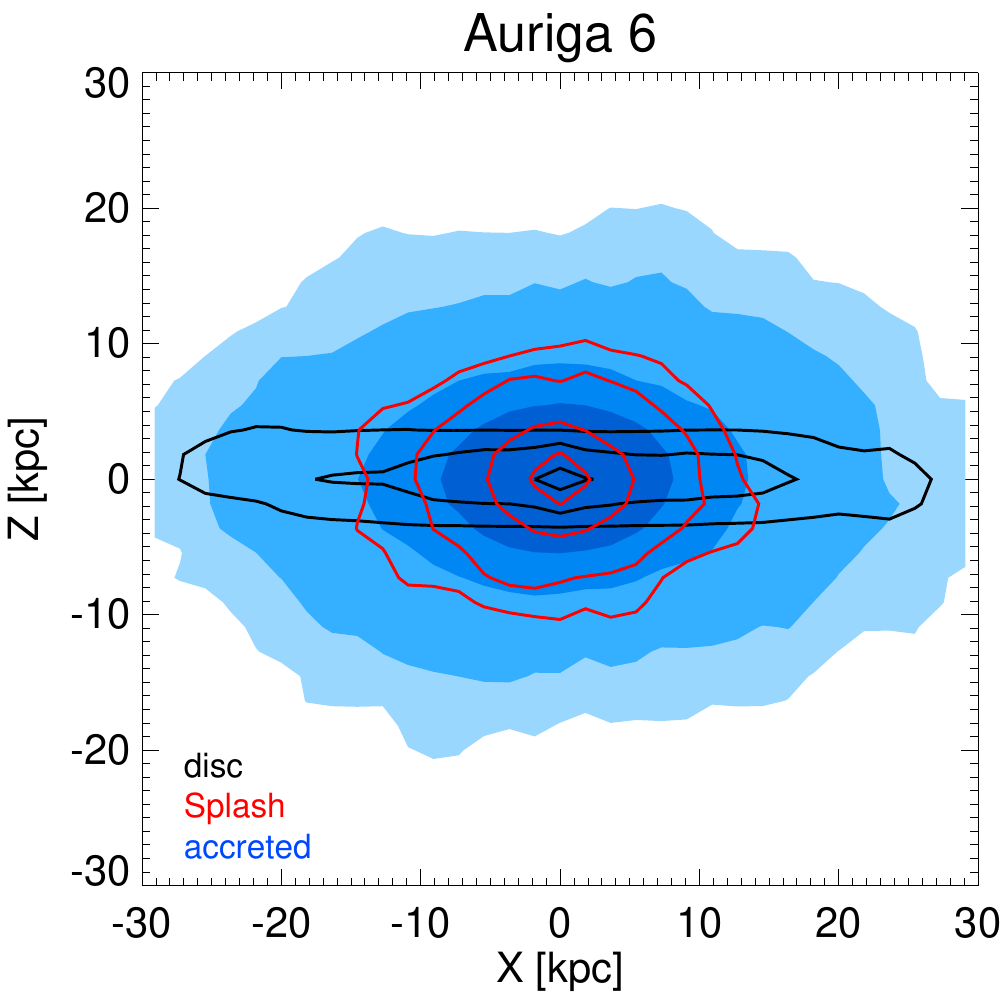}
  \includegraphics[width=0.32\textwidth]{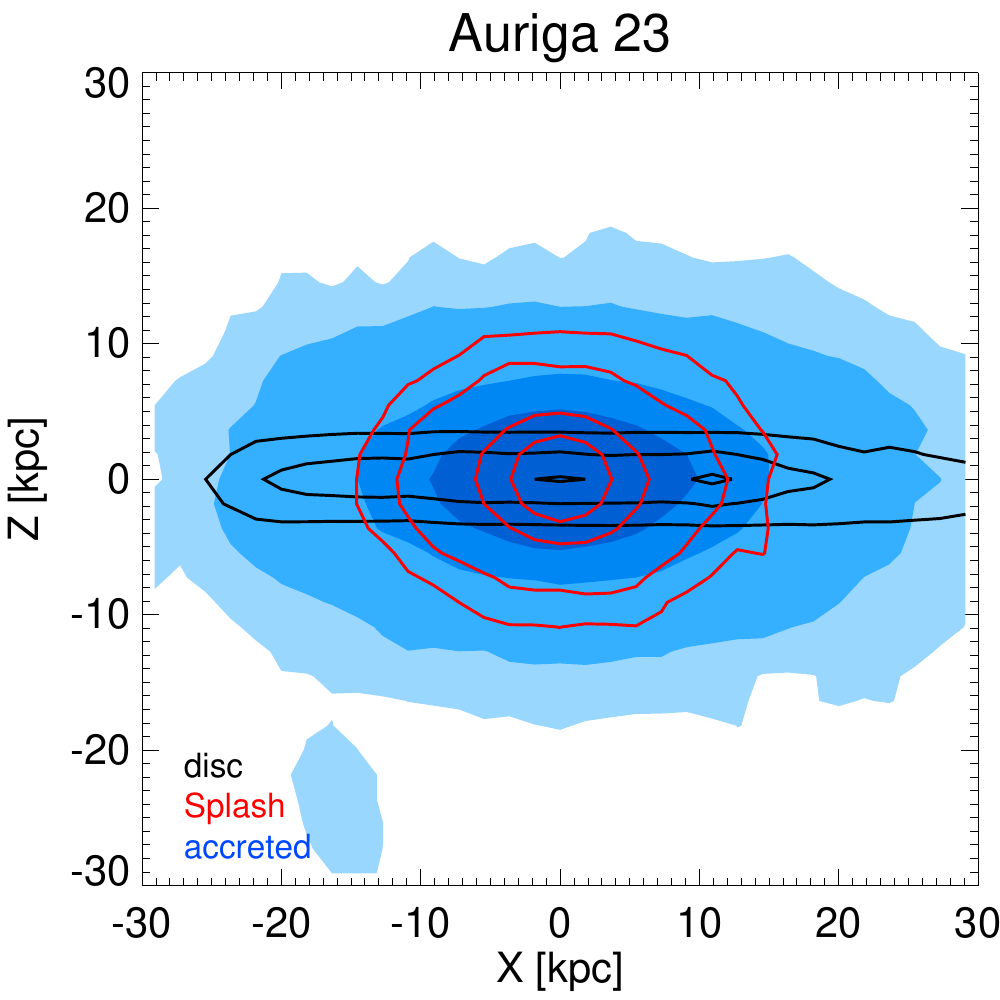}
  \includegraphics[width=0.32\textwidth]{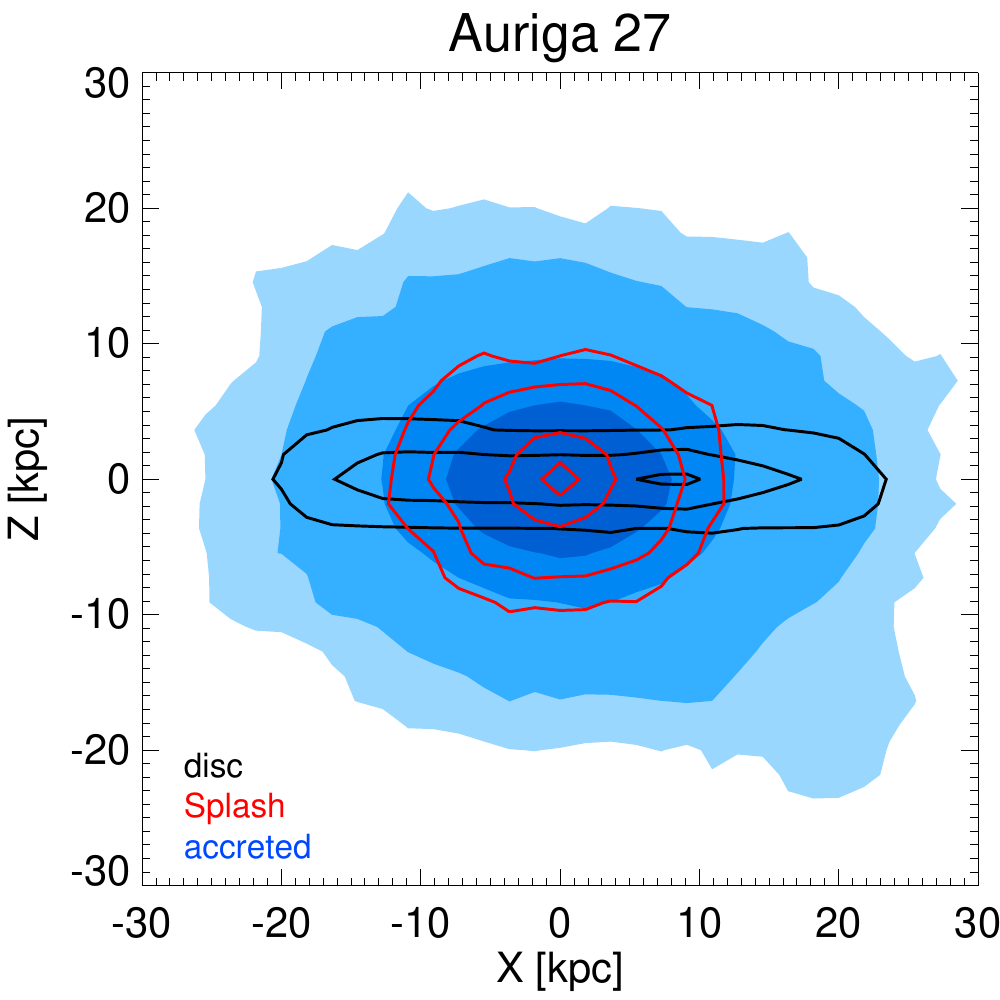}
  \caption[]{Present day density distribution in Auriga 6 (left), 23
    (middle) and 27 (right). Accreted (blue filled-in contours),
    in-situ after Splash (black) and Splash (red) populations are
    shown. The Splash stars are selected to have the age at least 0.5
    older than that marked by the vertical red line in
    Figure~\ref{fig:auexample} and $v_{\phi}<100$ km
    s$^{-1}$. Contours for both the accreted halo and the Splash
    populations are at the same density levels corresponding to the
    50th, 70th, 90th and 96th percentiles of the star count
    values. The disc contours are at the 92nd, 97th and 99.8
    percentiles of the disc density values}
   \label{fig:auden}
\end{figure*}

Here we tease out the details of the Splash formation from the
existing Cosmological simulations of Milky Way like galaxies.

\subsubsection{Auriga simulations}
The Auriga project \citep[][]{Grand2017} is one of the largest suites
of magneto-hydrodynamical high-resolution Galaxy simulations currently
available. Auriga uses the ‘zoom-in’ technique \citep[][]{Frenk1996,
  Jenkins2013} and the tree-PM moving-mesh code \texttt{AREPO}
\citep[see][]{Springel2010}. The subgrid physics model comprises a
spatially uniform photoionizing UV background, primordial and metal
line cooling, star formation, stellar evolution and supernovae
feedback, supermassive black hole growth and feedback, and magnetic
fields. A fixed set of chemical elements, including Fe and H, have
been consistently tracked within the simulations. Haloes and bound
(sub-)structures were identified using the FOF and SUBFIND algorithms
\citep[see][]{Davis1985, Springel2001}. MW analogues were selected
from the $100^3$ Mpc$^3$ DM-only periodic box of the EAGLE project
\citep[][]{Crain2015, Schaye2015}, based on their virial mass,
$\sim10^{12} M_{\odot}$, and an isolation criterion. The cosmological
parameters adopted for the simulations are those of Planck
Collaboration XVI \citep[][]{Planck2014}. In practice, we use the
publicly available Aurigaia mock catalogs described in
\citet{Aurigaia}, which provide access to the present day properties
of the stellar particles for six of the Auriga haloes which have been
simulated at the highest resolution level, (L3) with a baryonic mass
resolution of $\sim 10^4\,$M$_\odot$. These six haloes (no. 6, 16, 23,
24 and 27) are a subset of the original thirty haloes published in
\citet[][]{Grand2017}. We use the no-extinction, ICC branch of the
Aurigaia with the fiducial $30^{\circ}$ angle for the Solar
position. The ICC Aurigaia datasets are produced with the method of
\citet{Lowing2015} to generate mock stars from simulation star
particles. In our analysis, we include stars bound to the MW analogues
(not to the satellites) and decrease the metallicity ([Fe/H]) of stars
by a constant value of 0.5 dex in accordance with \citet{Fattahi2019}.

Figure~\ref{fig:auexample} shows the behaviour of the azimuthal
velocity as a function of metallicity (first and second column) and
age (third and fourth column) for stellar particles in Auriga 6, 23
and 27. Note that these galaxies are amongst the ones with a highly
radial stellar halo component as shown in
\citet{Fattahi2019}. Moreover, Auriga 23 is the clearest example of
the [$\alpha$/Fe] bimodality as described in \citet{Grand2018}. We
also checked the other three Aurigaia mocks and found the behaviour
consistent with what is presented below. To make comparison with the
data (as discussed in Section~\ref{sec:sample}) more appropriate, we
limit the stellar particles to those with $5 < R/{\rm kpc} < 11$ and
$0.5 <|z|/{\rm kpc} < 3$. The left column of the Figure shows the
row-normalized density of stars in the plane of $v_{\phi}$ and [Fe/H]
and can be compared directly with the observed distribution shown in
the third panel of Figure~\ref{fig:grid}. Each of the three Auriga
galaxies considered display a feature similar to the thin-thick disc
chevron at high [Fe/H] as well as the low angular momentum overdensity
at lower [Fe/H]. The next (second) column gives the distribution of
the median age in the plane of $v_{\phi}$ and [Fe/H]. As noted
earlier, the upper portion of the chevron is composed of the thin disc
stars, which is confirmed by the purple and dark blue colors
(i.e. typical ages $<6$ Gyr) in the region of the diagram. Moving
along the lower portion of the chevron, one encounters a strong age
gradient, supporting the inside-out thick disc growth hypothesis
discussed in \citet{Schonrich2017}. The metal-rich Splash stars with
low angular momentum can also be seen in the second panel in the case
of Auriga 23 and 27 where the last major merger happened some 10 Gyr
ago. In these two cases, at [Fe/H]$>-1$, the color changes from blue
to green to yellow and orange on moving from high positive $v_{\phi}$
to low and negative values. In Auriga 6, the last significant merger
happened later, $\sim$8 Gyr ago, thus the Splash has the median age of
the thick disc and the corresponding green color. Please note that in
the discussion above, instead of over-simplified ``the last major
merger'', a more appropriate nomenclature perhaps would be ``the last
major that reached the Solar neighborhood''. Interestingly, the end
phases of the mergers as measured here match well those reported in
\citet{Fattahi2019}: the differences are typically within 0.3 Gyr.

\begin{figure*}
  \centering
  \includegraphics[width=0.97\textwidth]{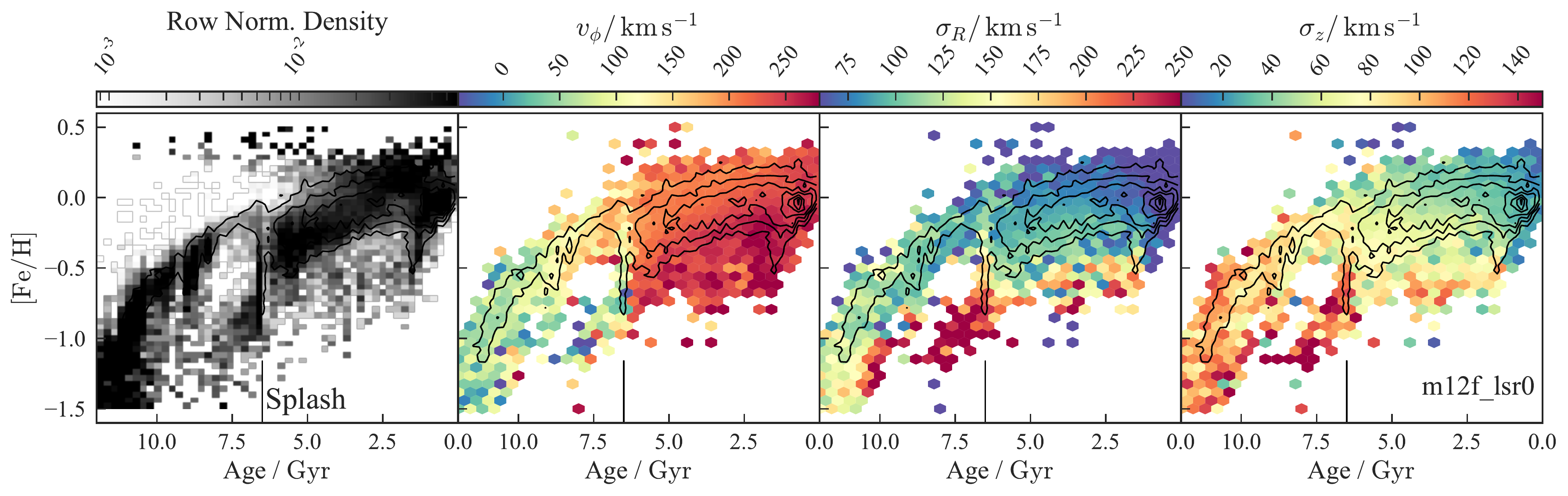}
  \includegraphics[width=0.97\textwidth]{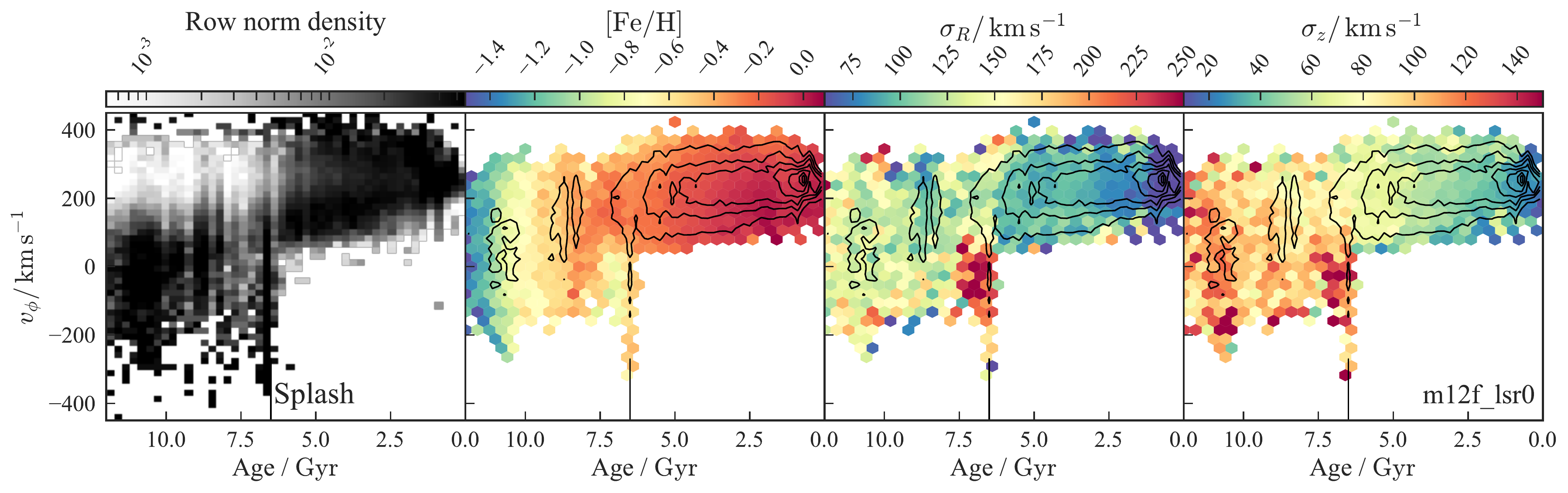}
  \caption[]{Solar neighbourhood stars selected from the \emph{ananke} Gaia mock for the Latte simulation m12f. The stars were selected to have parallaxes $>0.5\,\mathrm{mas}$, $|b|>15\,\mathrm{deg}$ and $G<17$. The top row of panels shows the row normalized density, mean azimuthal velocity, radial dispersion and vertical dispersion in bins of age vs. metallicity. The bottom row shows similar in the plane of age vs. azimuthal velocity. Note the Sausage-like merger event (marked with the vertical line) corresponding to the youngest metal-rich counter-rotating Splash stars, a subsequent increase in star formation and an increase in the radial velocity dispersion for stars older than the event. The contours are equally spaced in square-root of the number density.}
   \label{fig:latte}
\end{figure*}

The third column of Figure~\ref{fig:auexample} is the analog of the left
panel in Figure~\ref{fig:age} and shows the dependence of $v_{\phi}$
on stellar age for the metal-rich stars in the three Auriga
galaxies. The vertical Splash column is clearly visible in all cases
considered here. Vertical red line gives an approximate young edge of
the Splash and is calculated as the 20th percentile of the age
distribution of the metal-rich stars with $v_{\phi}<0$. Note that in
all examples, the ``thick'' (slow) disc population continues to form
after the major merger. In the fourth column, we split the
$v_{\phi}$-age distribution into accreted (blue) and in-situ (red)
components. Note that the edge of Splash (shown with the vertical red
line, see previous panel) coincides perfectly with the end of the
major accretion event in all three cases. We conjecture that i) the
youngest of the (local) retrograde in-situ and accreted stars
correspond to the final stages of a major merger event which ii) heats
up the pre-existing in-situ population kicking it onto highly
eccentric orbits.

In what follows we attempt to use the information in hand to deduce
the pre-Splash conditions indirectly (a more in-depth analysis of the
early state of the Auriga galaxies will be presented in Grand et al,
in prep). Accordingly, the horizontal red line marks the median
$v_{\phi}$ of the Splash stars, i.e. those to the left of the vertical
red line. In the case of Auriga 6 and 23 where the median azimuthal
velocity of the Splash stars is $100 < v_{\phi} < 200$, it is clear
that the galaxy was in a proto-disc state before the accretion as it
managed to retain much of its angular momentum. The case of Auriga 27
is less clear; this galaxy could have started in a spheroid
configuration which was further heated by the massive accretion
event. Finally, the fifth column of Figure~\ref{fig:auexample}
presents the age distributions for the accreted (blue), in-situ (red)
and Splash (black) stars (compare to the middle panel of
Figure~\ref{fig:age}). In all cases, the accreted population is
typically older than that of the Splash in agreement with the
observations presented above. In terms of the local star-formation
activity, the SF rate typically increases significantly during the
merger (continuous rise in the number of stars formed) and subsides
immediately after (as indicated by a reasonably flat distribution of
ages to the right of the vertical red line).

Figure~\ref{fig:auden} displays the present day density distributions
of the accreted (blue), and in-situ (red and black) populations in
Auriga 6 (left), 23 (middle) and 27 (right). To gain an uninterrupted
view of the Galaxy using the Aurigaia mock catalogs, we limit the
stellar tracers to those with absolute magnitude $M_V<-2$. In this
Figure, we split the in-situ population into two categories: the
Splash (red contours), i.e. stars born at least 0.5 Gyr before the end
of the major merger as indicated by the vertical red line in
Figure~\ref{fig:auexample} and those born after (black
contours). Structurally, Splash appears to be a smaller version of the
accreted halo, somewhat reminiscent of a fluffy classical bulge. It is
much rounder in shape compared to the rest of the in-situ stars, that
are mostly found at the low Galactic heights. In all three Auriga
examples, the Splash does not extend much beyond 10-15 kpc in both $R$
and $z$ dimensions. 

\begin{figure}
  \centering
  \includegraphics[width=0.97\columnwidth]{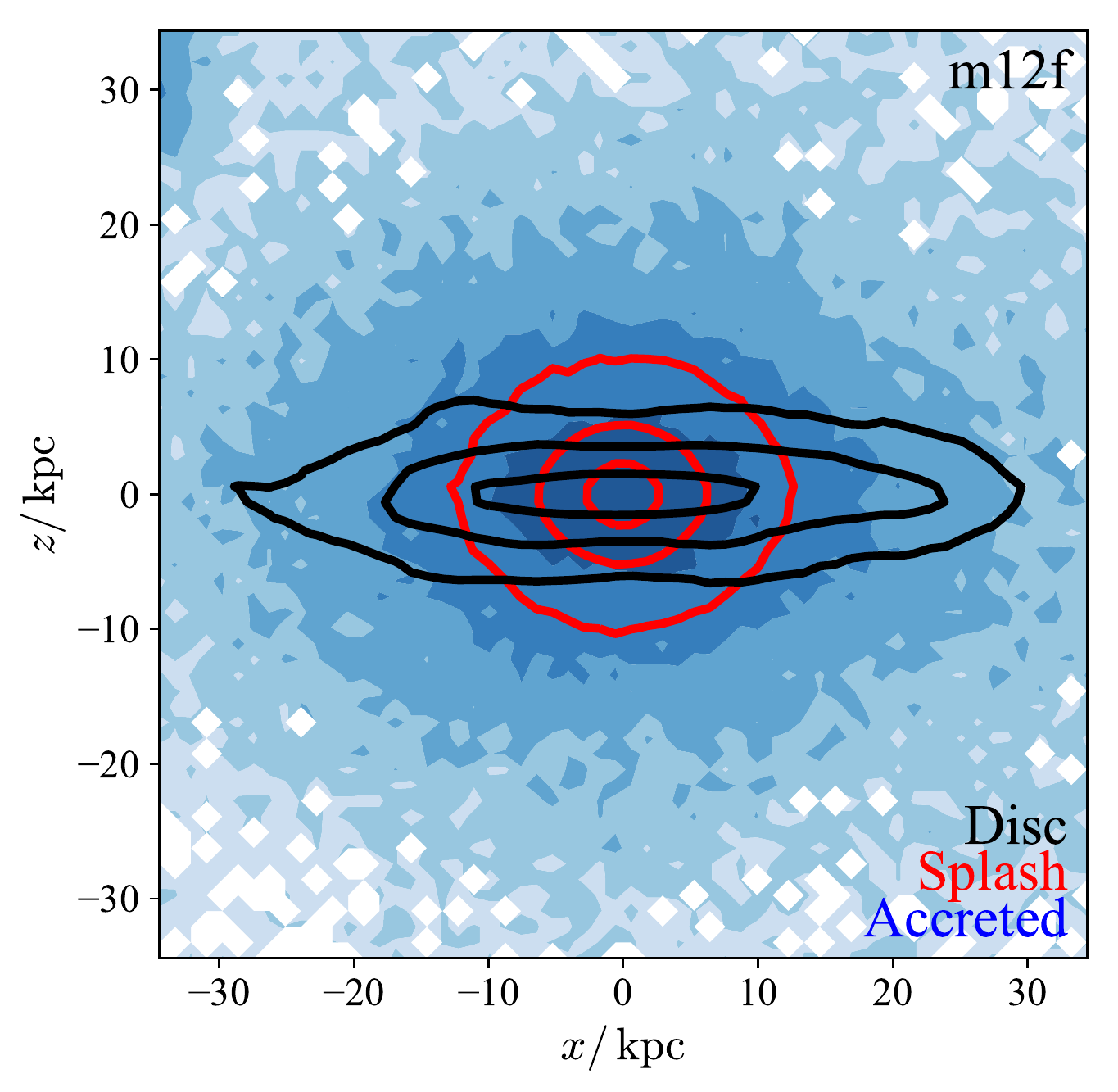}
  \caption[]{Present day density distribution for Latte simulation m12f (analogous to figure~\protect\ref{fig:auden} for the Auriga simulations). The disc component is metal-rich ($\mathrm{[Fe/H]}>-0.4$), rotating ($v_\phi>150\,\mathrm{km\,s}^{-1}$) and young (age<$6\,\mathrm{Gyr}$). The Splash and Accreted components are old (age>$7\,\mathrm{Gyr}$), retrograde ($v_\phi<0$) and split by tracks in age-metallicity space. We employ a further cut at age=$9.5\,\mathrm{Gyr}$ to define the accreted component so as to avoid contamination from in-situ stars. Contour levels are drawn at $0.1$, $1$ and $10\,\mathrm{per\,cent}$ of the peak density for the Splash and the disc, and additionally at $0.315$, $3.15$ and $31.5$ for the accreted component.}
   \label{fig:latte_density}
\end{figure}

\subsubsection{Latte simulations}

The Latte project \citep{Wetzel2016} is a suite of $\sim10$ hydrodynamical simulations of Milky Way mass galaxies. As with Auriga, the simulations use a zoom-in technique, selecting Milky Way analogues, based on their virial mass ($1-2\times10^{12}M_\odot$) and an isolation criterion, from a dark-matter-only periodic box of $85.5^3\,\mathrm{Mpc}^3$ in a $\Lambda$CDM cosmology. The simulations utilise the GIZMO code \citep{hopkins2015} which implements a mesh-free finite-mass method to solve the hydrodynamics and the tree-PM solver from GADGET-3 \citep{springel2005} for the gravity. The sub-grid physics model, FIRE-2  \citep{hopkins2018}, implements radiative heating and cooling including metal-line cooling tracking $11$ elements, a spatially-uniform UV background and a set of stellar feedback processes.

The \emph{ananke} project \citep{sanderson} used three of the Latte simulations (m12f, m12i, m12m) to generate mock Gaia observations of Milky Way-like simulations. This employed the resampling technique from \cite{sharma2011} coupled with a set of stellar isochrones \citep{Bressan2012}. The dust extinction is generated self-consistently from the gas in the simulation assuming dust traces metal-rich gas. For each simulation, three different solar locations are employed. We have extracted a solar neighbourhood sample of $100,000$ stars from the \emph{ananke} mock Gaia catalogue for the m12f simulation (using the Sun location lsr0) subject to the selection: parallax$>0.5\,\mathrm{mas}$, $|b|>15\,\mathrm{deg}$ and $G<17$, which approximately resembles the selection from a typical spectroscopic survey. Of the three simulations, m12f \citep[first presented in][]{garrisonkimmel2017} was selected due to its resemblance to the Milky Way. For example, the stars exhibit a bimodal $[\alpha/\mathrm{Fe}]$-$[\mathrm{Fe}/\mathrm{H}]$ distribution.

In Fig.~\ref{fig:latte} we show the distributions of our selected
sample in the spaces of age vs. metallicity and age vs. azimuthal
velocity. We observe that the Milky Way chemical evolution proceeds as
expected up to a metallicity of $\sim-0.3\,\mathrm{dex}$ at
$6.5\,\mathrm{Gyr}$. At this point a merger event (shown clearly by
the hotter $\sigma_R$ and $\sigma_z$ kinematics and lower chemical
evolution track) occurs (there is an additional smaller merger event at $\sim10.5\,\mathrm{Gyr}$). This merger epoch corresponds to the youngest
ages for counter-rotating metal-rich ([Fe/H] $>-0.4$) stars, coincides
with a slight resetting of the metallicity track (upper left panel) as
metal-poor accreted gas dilutes the Milky Way gas and signifies the
point at which the radial dispersion suddenly drops \citep[third lower
  panel but see also see Fig. 3 of][]{sanderson}. It is this feature
of m12f that gives rise to the bimodal
$[\alpha/\mathrm{Fe}]$-$[\mathrm{Fe}/\mathrm{H}]$ distribution. After
this merger epoch, star formation accelerates (note the rise in the
black density contours), likely prompted by the merger event, and disc
evolution proceeds in a more regular fashion with essentially no
counter-rotating stars. The chevron structure of the thin and thick
discs is visible after the merger demonstrating that thick disc-like
evolution can occur \emph{after} the merger event (even for stars
formed long after the merger, a bimodal chemical distribution
persists).  Quantitatively, the dispersions of the different
populations correspond nicely to the figures in
Table~\ref{tab:comp}. The young thin disc stars (defined as [Fe/H]
$>-0.4$, age $<1.5\,\mathrm{Gyr}$ and
$v_\phi>200\,\mathrm{km\,s}^{-1}$ -- corresponding almost exclusively
to the low alpha sequence) have dispersions in $R$ and $z$ of
$(61,30)\,\mathrm{km\,s}^{-1}$ respectively. The thick disc (defined
as [Fe/H] $>-0.4$, $4.5<$ age $/\,\mathrm{Gyr}<6.5$ and
$100<v_\phi<200\,\mathrm{km\,s}^{-1}$ -- corresponding to the high
alpha sequence) has $(93,66)$. The Sausage-like merger remnant
(defined as [Fe/H] $<-0.4$, $6.5<$ age $/\,\mathrm{Gyr}<9.5$ and
$v_\phi<0$) has $(219, 129)$ and finally the Splash-like Milky-Way
component (defined as [Fe/H] $>-0.4$, age $>6.5\mathrm{Gyr}$ and
$v_\phi<0$) has $(153, 101)$. In general, the simulation is
kinematically hotter than the Milky Way \citep{sanderson}, but the
hierarchy and respective ratios show good correspondence with the
observations. Finally, we present the density distributions of the different components in Fig.~\ref{fig:latte_density}. The distribution is very similar to that observed in the analogous Auriga plots in figure~\ref{fig:auden}. Here we define the components in a similar way as for the local sample but requiring the disc has age$<6\,\mathrm{Gyr}$ and $v_\phi>150\,\mathrm{km\,s}^{-1}$, and using a by-eye chemical evolution track in age-metallicity space to separate out the in-situ from accreted halo. We observe that the Splash component is spheroidal with a similar flattening to the accreted component. Without full histories of the star particles we cannot
state definitively that the simulation reflects the Splash hypothesis
presented in this paper. However, it seems highly likely. An
alternative, more unlikely hypothesis is that the proto-disc was near
spheroidal and after the merger a much colder gas disc formed.

\subsection{Virial Considerations}
\label{sec:virial}

Finally, we can cross-check our ideas by using the virial theorem to estimate the mass $\Ms$ of the infalling satellite. As the satellite falls a distance $\Delta r$, then it loses energy
\begin{equation}
\Delta E \approx -\Ms {d\phi\over dr} \Delta r \approx -{\Ms
v_0^2\over r} \Delta r
\label{eq:one}
\end{equation}
where $\phi$ is the total potential, and the Galaxy is assumed to have
a flat rotation curve $v_0$. The energy loss of the satellite under
dynamical friction is pumped into heating both the proto-disc and the
halo. As the Sausage encounter is low inclination, we assume that most
of the heating takes place in the proto-disc to produce both the thick
disc and the Splash.

Some of the energy deposited in the proto-disc is converted to potential energy, as the disc thickens in response to the vertical heating. We use the virial theorem (per unit surface area of the disc) to estimate how $\Delta E$ turns into kinetic and potential energy. The vertical component reads 
\begin{equation}
2T + W \approx 0,
\end{equation}
where we have discarded the time-dependent term as we are assuming the system settles into equilibrium after the encounter has ended. If $H$ is the scaleheight of the proto-disc and $\Sigma$ the surface density, then the kinetic and potential energy (per unit surface area) are
\begin{equation}
T \approx \frac{1}{2}\Sigma \sigma_z^2 \qquad\qquad
    W_1 \approx -\pi G \Sigma^2 H
\end{equation}
 The proto-disc has a total potential energy $W = W_1 + W_2$ because it resides in the gravitational potential of itself and the dark halo, with the latter term for a flat rotation curve galaxy being
\begin{equation}
    W_2 \approx -\frac{3}{2}\Sigma H^2 G {M_h(r) \over r^3}.
\end{equation}    
The total energy change $\Delta E = \Delta T + \Delta W$, and so using the virial theorem
\begin{equation}
  {1\over 2} \times  {\Delta E \over 2 \pi r \Delta r} \approx \frac{\pi}{2} G \Sigma^2 \Delta H \left( 1 +  3{H M_h(r)\over \pi r^3 \Sigma}\right)
    \label{eq:two}
\end{equation}
where we have assumed that only half the energy change goes into vertical heating (and the other half into radial heating). Inserting eq~(\ref{eq:one}), we obtain for the change in scaleheight of the proto-disc
\begin{equation}
    \Delta H \approx {\Ms v_0^2\over 2\pi^2 G \Sigma^2 r^2},
\end{equation}
where we have omitted the small correction term in brackets in eq.~(\ref{eq:two}), which is of the order 15 \% This gives the change in vertical velocity dispersion of the proto-disc as
\begin{equation}
    \Delta \sigma_z^2 \approx {\Ms v_0^2 \over 2\pi \Sigma r^2}
\end{equation}
We now assume that the proto-disk density law $\Sigma$ is of Mestel form and that it provides half the contribution to $v_0$ locally, with the dark halo providing the rest. Normalising to the Solar neighbourhood, we obtain
\begin{equation}
    \Delta \sigma_z^2 \approx  0.25 v_0^2 \left( {\Ms \over M_{\rm disk}} \right) \left( R_\odot \over r \right).
\end{equation}
where $M_{\rm disk}$ is the mass contained within $R_\odot$. Referring to the velocity dispersions in Table 1, we now assume that the encounter excited both the stars now in the Splash and the thick disc. This gives $\Delta \sigma_z^2 \approx 2700$ km$^2$s$^{-2}$ (using the data in Table 1 and weighting the thick disc and the Splash by their fractional contribution)

Assuming that the amplitude of the flat rotation curve is $v_0 = 230$ kms$^{-1}$, this finally gives
\begin{equation}
    \left( {\Ms \over M_{\rm disk}} \right) \approx 0.20.
\end{equation}
On energetic grounds, the Sausage is able to excite stars vertically
into thick disc and Splash orbits if it is roughly an order of
magnitude less massive than the proto-disc.

\subsection{What's in the Splash?}
\label{sec:what}

As shown clearly from Figure~\ref{fig:age}, the bulk of the MW stars with
ages similar to that of the Splash carry plenty of angular momentum
today, implying that before the massive ancient merger the Galaxy was
likely a proto-disc. The most direct interpretation of this
observation (also supported by our analysis of the hydro-dynamical
zoom-in simulations discussed above; see Figures~\ref{fig:auexample}
and~\ref{fig:latte}) is that the Splash is composed of the stars
dispersed from the proto-disc of the Galaxy during the accretion event
which created the {\it Gaia} Sausage. In this regard, of the many
scenarios of in-situ halo formation proposed in the literature,
the heated in-situ channel -- as discussed in \citet{McCarthy2012} --
appears to be the most viable, at least here, in the Milky Way. In
this picture, the Splash may simply represent the most energetic portion
of the thick disc. The seemingly smooth transition from the Splash to
the thick disc as observed in the many panels of Figure~\ref{fig:grid}
would argue in favour of this interpretation. On the other hand,
the Splash appears to be an additional component in the (albeit rather
simplified) azimuthal velocity decomposition presented in
Figure~\ref{fig:vel_fit}. Another important difference between the two components
is that, according to Figures~\ref{fig:age}, ~\ref{fig:auexample} and~\ref{fig:latte},
the Splash formed only during the last major merger, while the thick disc
appears to have formed before, during and after the event. For
example, the left panel of Figure~\ref{fig:age} shows that the thick
disc formation likely continued until 8 Gyr ago, i.e. 1-2 Gyr after
the Splash, in good agreement with previous studies \citep[see
  e.g.][]{Haywood2013}.

Note however that given the data in hand, it is impossible to
completely rule out other (perhaps more exotic) modes of the Splash's
formation. Apart from the possibilities already discussed in
\citet{Cooper2015}, let us point out two additional channels: first,
given the relatively early time of the {\it Gaia} Sausage merger, it
is quite likely that in the accretion event, in addition to the dark
matter and stars, large amounts of unused gas were also dumped onto
the nascent Milky May. In such a gas-rich merger, the violent
interaction between the gaseous discs of the two galaxies could toss
the gas clouds around, causing some of the promoted star-formation to
occur on rather eccentric, halo-like orbits. Another Splash formation
mechanism can operate within gaseous outflows launched as a result of
the profuse galactic feedback, induced either by star-formation or AGN
activity. \citet{Maiolino2017} present the first unambiguous example
of ongoing star-formation within a galactic outflow. Subsequently,
\citet{Gallagher2019} demonstrate that at least $\sim30\%$ of all
outflows in the MaNGA DR2 samples exhibit clear signs of
star-formation, in agreement with the independent analysis of the
MaNGA data by \citet{DelPino2019}. These striking discoveries have
been preceded by a decade of mounting evidence for molecular gas
outflows \citep[see
  e.g.][]{Feruglio2010,Sturm2011,Cicone2014}. Furthermore, a number of
independent analytical and numerical models now exist that
substantiate the possibility of star formation on highly radial orbits
within galactic outflows
\citep[e.g.][]{Ishibashi2012,Zubovas2013,Ishibashi2013,Zubovas2014,Wang2018,Deca2019}. These
findings imply that star-formation activity within outflows is
ubiquitous; as noted by \citet{Gallagher2019}, regular isolated disc
galaxies bear signs of this process. As is clear from the right column
of Figure~\ref{fig:auexample}, the SF rate increases sharply during
the last major merger, thus creating conditions favourable for the
production of the Splash population inside the feedback-induced
out-flowing gas. \citet{Maiolino2017} and \citet{Gallagher2019}
discuss the fate of the stars born this way. They note that it is
currently impossible to have a meaningful constraint on the orbital
properties of the outflow-born stars. However, they conjecture that as
soon as the stars are born, they decouple from the outflow and
continue their motion in the host's potential ballistically. They
estimate that the bulk of this population will be limited to the inner
few kpc of the Galaxy in good agreement with our measurements of the
extent of the Splash. We envisage that future studies of the chemical
abundance gradients across the Splash may shed light on the genesis of
this population.

\subsection{Conclusions}

We have analyzed orbits, ages and alpha-abundances of a large sample
of nearby stars, focusing on the metal-rich, $-0.7<$[Fe/H]$<-0.2$
population of stars with low angular momentum, which we name the
Splash. Our principal findings are summarised below.

\begin{itemize}

\item In terms of metallicity, the Splash can be seen most easily
  between [Fe/H]=-0.7 and [Fe/H]=-0.2. More metal-poor stars typically
  have more eccentric orbits and larger vertical velocity dispersion,
  while more metal-rich stars have higher angular momentum (see middle
  row of Figure~\ref{fig:grid}). Note however that we do not imply
  that the Splash population is limited to this range of iron
  abundance and believe that it likely stretches to lower values of
  [Fe/H], where it is overwhelmed by the accreted stars. Recently,
  \citet{Sestito2019} detected a significant number of ultra-metal
  poor stars wit orbits confined to the disc plane. These stars
  could well be the denizens of the Galaxy's proto-disc which was
  heated by massive accretion events (in line with the conclusions
  reached in this paper). Their result also implies that there exists
  a population of ancient stars whose orbits had not been much
  affected by the cataclysmic events that lead to the creation of the
  {\it Gaia} Sausage. This can be explained if not all of the
  proto-disc was destroyed. In that case however, it is difficult to
  imagine how stars at large distances from the Galactic centre could
  remain unaffected. Perhaps instead, these stars were born in a
  disc-like configuration during or immediately after the last major
  merger. Alternatively, it is easier to preserve the Galactic
  proto-disc, if the Splash was created via a process that does not
  require the disc destruction (see e.g. Section~\ref{sec:what}). Finally, these stars may have been `protected' in the inner Galaxy during the merger event, and then migrated outwards. This scenario requires efficient radial migration in the (potentially narrow) window between the merger event and the Galactic bar forming and subsequently trapping stars in the inner Galaxy.

\item We show that in the $v_{\phi}$-[Fe/H] plane there exists a
  chevron-like feature, resembling a $>$ sign and corresponding to the
  ``thin-thick'' disc bifurcation. The Splash is located at and below
  the break in the $v_{\phi}$-[Fe/H] sequence of the ``thick'' disc
  and has metallicity typical of both the ``thick'' and the ``thin''
  disc populations (third panel of the top row of
  Figure~\ref{fig:grid}). In fact, in many other dimensions studied
  here, the Splash connects smoothly to the ``thick'' disc,
  e.g. azimuthal velocity, alpha abundances and ages. The Splash
  properties however are extreme compared to the bulk of the ``thick''
  disc. It has little angular momentum and plenty of retrograde stars,
  with most eccentricities in excess of $e>0.5$, rather unusual for
  the disc. At a given metallicity, the Splash's alpha-abundances are
  typically the highest attained by the ``thick'' disc stars (see
  bottom row of Figure~\ref{fig:grid}) while its ages are the
  oldest. It is therefore difficult not to conclude \citep[in
    agreement with][]{Bonaca2017,Haywood2018,DiMatteo2018,Gallart2019}
  that the Splash is the ancient portion of the ``thick'' or proto-
  disc of the Galaxy.

\item The cleanest Splash selection is that delivered by two simple
  cuts: $v_{\phi}<0$ and [Fe/H]$>-0.7$. This is the portion of the
  parameter space that suffers the lowest contamination and is where
  Amarante et al. (2019) detect Splash-like material (see their fig
  7). Note however that according to our kinematic modelling presented
  in Section~\ref{sec:sauspl}, the Splash population reaches to much
  higher values of $v_{\phi}$ and has a clear net positive spin. Our
  3-component Gaussian model is rather naive, and should not be
  over-interpreted. However, it does show quite clearly both the
  necessity for an additional kinematic component (the knee at around
  $v_{\phi}\sim0$ km s$^{-1}$) and the extent of this component to
  $v_{\phi}$ as high as 100-200 km s$^{-1}$. In connection to this,
  the recently discovered giant, prograde and metal-rich stream Nyx
  \citep[see][]{Nyx} may well be nothing but a piece of the
  Splash. Additionally, it is interesting to point out that our
  high-metallicity boundary for the Splash population of $<-0.2$
  \citep[also see][]{DiMatteo2018} coincides with the value of [Fe/H]
  where changes in the disc's chemo-dynamical properties had been
  noted before. For example, this is the highest metallicity where the
  bimodality in [$\alpha$/Fe] can be established \citep[see
    e.g.][]{Nidever2014}. This is also the boundary chosen by
  \citet{Snaith2015} to demarcate the inner and outer thin disc
  populations.
  
\item The Splash can be seen most clearly in Figure~\ref{fig:age} where it
  is visible as a vertical band of stars with a large range of
  $v_{\phi}$ limited to a constant and narrow range of ages ($<10$
  Gyr). We point out a strikingly synchronized truncation in the age
  distribution of the retrograde metal-poor (likely accreted) stars
  and the retrograde metal-rich (likely in-situ) stars (see middle and
  right panels of Figure~\ref{fig:age}). In line with
  \citet{Gallart2019} we argue that the youngest age of the Splash
  stars can be used to date the last major merger experienced by the
  Galaxy, i.e the accretion that lead to the creation of the Gaia
  Sausage. Note, however, that we disagree with the conclusion of
  \citet{Gallart2019} that the star-formation histories of the halo
  and the Splash are indistinguishable. As the middle panel of our
  Figure~\ref{fig:age} clearly shows, the age distribution of the
  Splash stars peak some $\sim1$ Gyr later. This is consistent with
  both i) the fact that the Splash stars are significantly more
  metal-rich than those in the halo and ii) the hypothesis that the Splash
  may originate from the proto-disc, which would also require some
  time to assemble.

\item We have also attempted to map the 3D structure of the Splash
  well beyond the Solar neighborhood. We have obtained consistent
  constraints on the spatial extent and the shape of the Splash
  population using two independent datasets, that of
  \citet{SandersDas2018} and the SDSS \citet{Xue2014} and LAMOST
  K-giants. While near the Sun, there are as many metal-rich stars
  with $v_{\phi}<100$ km s$^{-1}$ as there are metal-poor ones, the
  ratio of metal-rich to metal-poor low-angular momentum stars drops
  quickly with Galacto-centric radial distance and height. In
  particular, at $R>15$ kpc and $|z|>10$ kpc, this ratio can be
  $\sim5\%$ or lower (see Figures~\ref{fig:ratio}, ~\ref{fig:kgrid}
  and ~\ref{fig:kg_lamost_spatial}). The minuscule size of the Splash
  compared to the rest of the Galactic stellar halo or even the Gaia
  Sausage makes sense if the Splash was born out of the proto-disc of
  the Milky Way, itself believed to be rather small. Interestingly,
  the distribution of the oldest Galactic long-period variables also
  exhibits a sharp truncation around 15-20 kpc \citep[see Figure 13
    of][]{Grady2019}. \citet{EmmaHaloTail} use APOGEE spectroscopy and
  Gaia DR2 astrometry to identify three distinct groups of metal-rich
  [Fe/H]$>-0.75$ stars at Galactic heights of $|z|>10$ kpc. They show
  that of these three, one can be associated with the Sgr stream (also
  see Section~\ref{sec:kg} of this paper) and one with the recently
  kicked-up disc stars in the anti-centre region. The third and final
  group has a low-amplitude prograde rotation (decreasing with $|z|$)
  and high values of [Mg/Fe]. It is almost certain that this third
  group detected by \citet{EmmaHaloTail} is part of the Splash.
  
\item We have inspected examples from the Auriga and Latte suites of
  Cosmological zoom-in simulations of the Milky Way's formation,
  finding that example galaxies in both suites produce a Splash-like
  structure as a result of the interaction of a massive dwarf galaxy
  with the pre-existing in-situ (typically disc-like) population
  \citep[in agreement with][]{Bonaca2017, Monachesi2019}. We show that
  the epoch of this major merger can be accurately calculated via two
  different routes: either relying on the youngest ages of the
  accreted stars observed locally at redshift $z=0$ or using the ages
  of the metal-rich in-situ stars on retrograde orbits. We find strong
  but indirect evidence that in several of the inspected Auriga
  simulations a proto-disc is transformed into a more spheroidal
  Splash component by a massive merger event. All of the 6 Aurigaia
  galaxies analysed here have a detectable Splash population of
  varying significance, demonstrating how easy it is to kick up
  in-situ stars onto halo-like orbits in agreement with
  \citet{JB2017}. We therefore propose to exploit the intimate
  connection between the last major merger and the Splash to calculate
  the mass of the in-falling satellite as well as the geometry and the
  kinematics of the merger (we describe a simple analytic attempt in
  Section~\ref{sec:virial}). The details of the interaction between
  the {\it Gaia} Sausage progenitor and the in-situ population as well
  as the ensuing transformation of the Galactic proto-disc will be
  discussed in the upcoming publication (Grand et al, in prep).
\end{itemize}  

The picture of the early days of the Milky Way as painted by the
observations and the simulations discussed here is complex. There is
evidence that the influence of the last massive accretion event may
well have been both destructive and constructive. The ancient head-on
collision wrought havoc in the young Galaxy, heating, scrambling and
possibly truncating some of its disc. Yet it may have also helped to
promote star-formation and delivered fresh gas supplies to be used in
the subsequent disc growth. The Milky Way never looked the same after
the Biggest Splash.

\section*{Acknowledgments}

The authors wish to thank GyuChul Myeong, Emma Fern{\'a}ndez-Alvar,
Marie Martig, Sergey Koposov, Ted Mackereth, Daisuke Kawata, Roberto
Maiolino, Sergey Khoperskov and Francesca Fragkoudi for many useful
discussions that helped to improve this manuscript. Special thanks go
to Robyn Sanderson and Andrew Wetzel for their help with the Latte
simulations.

This research made use of data from the European Space Agency mission
Gaia (http://www.cosmos.esa.int/gaia), processed by the Gaia Data
Processing and Analysis Consortium (DPAC,
http://www.cosmos.esa.int/web/gaia/dpac/consortium). Funding for the
DPAC has been provided by national institutions, in particular the
institutions participating in the Gaia Multilateral Agreement. This
paper made used of the Whole Sky Database (wsdb) created by Sergey
Koposov and maintained at the Institute of Astronomy, Cambridge with
financial support from the Science \& Technology Facilities Council
(STFC) and the European Research Council (ERC).  The Auriga simulations used the DiRAC Data Centric system at Durham 
University, operated by the Institute for Computational Cosmology on 
behalf of the STFC DiRAC HPC Facility (\url{www.dirac.ac.uk}). This 
equipment was funded by BIS National E-infrastructure capital grant 
ST/K00042X/1, STFC capital grant ST/H008519/1, and STFC DiRAC Operations 
grant ST/K003267/1 and Durham University. DiRAC is part of the National 
E-Infrastructure. 

M.C.S acknowledges support from the National Key Basic Research and
Development Program of China (No.2018YFA0404501) and NSFC grant
11673083. AJD is supported by a Royal Society University Research
Fellowship, and AF by a European Union COFUND/Durham Junior Research Fellowship (under EU grant agreement no. 609412).

\bibliography{references}

\label{lastpage}

\end{document}